\documentclass[11pt]{article}
\usepackage{epsfig} 
\usepackage{graphics}
\newcommand{\sect}[1]{ \section{#1} \setcounter{equation}{0} }

\newcommand{\xslash}{x \! \! \! /}
\newcommand{\partialslash}{\partial \! \! \! /}
\newcommand{\half}{\mbox{\small{$\frac{1}{2}$}}}

\newcommand{\threehalves}{\mbox{\small{$\frac{3}{2}$}}} 
\newcommand{\MSbar}{\overline{\mbox{MS}}} 
\newcommand{\NA}{N_{\!A}}
\newcommand{\Nc}{N_{\!c}}

\begin{document}

\title{Critical exponent $\eta$ at $O(1/N^3)$ in the chiral XY model using the
large $N$ conformal bootstrap}
\author{J.A. Gracey, \\ Theoretical Physics Division, \\ 
Department of Mathematical Sciences, \\ University of Liverpool, \\ P.O. Box 
147, \\ Liverpool, \\ L69 3BX, \\ United Kingdom.} 
\date{}
\maketitle 

\vspace{5cm} 
\noindent 
{\bf Abstract.} We compute the $O(1/N^3)$ correction to the critical exponent
$\eta$ in the chiral XY or chiral Gross-Neveu model in $d$-dimensions. As the
leading order vertex anomalous dimension vanishes, the direct application of 
the large $N$ conformal bootstrap formalism is not immediately possible. To
circumvent this we consider the more general Nambu-Jona-Lasinio model for a
general non-abelian Lie group. Taking the abelian limit of the exponents of 
this model produces those of the chiral XY model. Subsequently we provide
improved estimates for $\eta$ in the three dimensional chiral XY model for 
various values of $N$. 

\vspace{-17.5cm}
\hspace{13.2cm}
{\bf LTH 1251}

\newpage 

\sect{Introduction.}

The Gross-Neveu model, \cite{1,2}, is one of the core quantum field theories 
that is on a similar level to scalar $\phi^4$ theory since it too has a quartic
self-interaction but is purely fermionic. By contrast although it is 
renormalizable in two space-time dimensions rather than four, more 
interestingly it is asymptotically free, \cite{1}. It shares this feature with 
four dimensional non-abelian gauge theories such as Quantum Chromodynamics 
(QCD) \cite{3,4} and, moreover, has the property of dynamical symmetry breaking
leading to non-perturbative mass generation, \cite{1}. In more recent years the
original Gross-Neveu model of \cite{1} and its extensions have been studied in 
the context of ultraviolet completion, \cite{5,6}. This is where the two 
dimensional theory is extended to become a renormalizable one in four 
dimensions which is achieved in several stages. The first is to introduce a 
Hubbard-Stratonovich transformation which produces an auxiliary scalar field. 
In two dimensions this field becomes dynamical and corresponds to a bound state
of two fermions, \cite{1}. In the ultraviolet completion to four dimensions, 
however, this auxiliary field develops a fundamental propagator, \cite{5}. So
to ensure four dimensional renormalizability it is necessary for the scalar 
field to have a quartic self-interaction. This in essence describes the 
structure of a Gross-Neveu-Yukawa theory which is the ultimate point of the 
ultraviolet completion exercise, \cite{5}. Indeed this particular theory has
been of interest for many years due to its potential to model a composite Higgs
field in extensions of the Standard Model, \cite{7}, for example.

Our description of this connection between two and four dimensions is primarily
for the original Gross-Neveu model which has a discrete chiral symmetry. In 
this context it is sometimes referred to as the Ising Gross-Neveu model. It is 
not the only version of a Gross-Neveu model in that alternatively one can endow
the basic quartic fermion interaction with a continuous chiral symmetry. In two
dimensions this is known as the chiral Gross-Neveu model, \cite{1}, and that 
has an ultraviolet completion to what is termed the chiral XY model, \cite{6}. 
This model as well as the Ising case have been the subject of considerable 
interest in the last decade or so due to its underlying connection with models 
of phase transitions in graphene. Stretching a sheet of carbon atoms can change
the electrical properties of the material from a semi-conductor to a Mott 
insulator, \cite{8,9}. Indeed the problem of evaluating the critical exponents 
describing the phase transition in the dimension of interest, which is three, 
has generated an immense amount of activity. The main theoretical approaches 
include the application of the functional renormalization group, \cite{10,11},
conformal bootstrap, \cite{12,13}, Monte Carlo methods, \cite{14}, the 
$\epsilon$ expansion, \cite{15,16}, and the large $N$ expansion, \cite{17}. We 
note that this is a representative set of recent articles rather than a 
definitive one. The connection between the $\epsilon$ expansion and the large 
$N$ approach is quite close. For the former the renormalization group functions
are calculated to high loop order in the ultraviolet completed four dimensional 
model. Then the critical exponents at the $d$~$=$~$4$~$-$~$2\epsilon$ 
dimensional Wilson-Fisher fixed point are computed before the $\epsilon$ 
expansion of an exponent is summed up prior to setting $\epsilon$~$=$~$\half$. 
In some Gross-Neveu universality classes the same exercise has been 
additionally performed in the original two dimensional theory. In that case the
exponents are determined at the Wilson-Fisher fixed point in an 
$\bar{\epsilon}$ expansion near $d$~$=$~$2$~$+$~$2\bar{\epsilon}$ dimensions 
even though the theory is only perturbatively renormalizable in strictly two 
dimensions. Despite these respective theories not being in the neighbourhood of
three dimensions, accurate estimates for exponents have been extracted in 
certain cases. For instance, the current situation in the Ising Gross-Neveu 
model has been summarized in \cite{16}. 

By contrast in the large $N$ critical point approach pioneered in scalar $O(N)$
models in \cite{18,19,20}, the first three terms of critical exponents have 
been computed in $d$-dimensional space-time as a function of $d$. The large $N$
expansion is renormalizable, \cite{21,22}, primarily due to $1/N$ being a 
dimensionless parameter. Using the first three terms in the series for 
relatively low $N$, estimates of exponents in three dimensions have been shown 
to be competitive with those of the other techniques mentioned. It is these
large $N$ exponents depending on $d$ that drive the close connection with the
exponents determined from the $\epsilon$ expansion of the explicit perturbative
renormalization group functions. This is due to the general observation that
the exponents depend on both variables $N$ and $\epsilon$. So expanding those 
in powers of $1/N$ and $\epsilon$ the expressions derived using either 
technique separately should be in agreement in the area where terms in the 
double expansion overlap. Indeed this is the case in many models. Moreover the 
large $N$ exponents can be expanded to all orders in $\epsilon$ at each order 
in $1/N$ thereby providing non-trivial checks on future perturbative results.

While the main focus so far has been on the Ising and chiral Gross-Neveu models
and their connections with graphene phase transitions, the chiral XY model
itself has been explored in relation to certain phases in matter. For instance,
the correlated symmetry protected phases were investigated in depth in 
\cite{23}. As this model clearly is of importance to these areas it is worth 
summarizing the current status of the analytic evaluation of the exponents. For
instance, the renormalization group functions of the chiral XY model are 
available to four loops, \cite{6,15}. Equally the $\epsilon$ expansion of the 
large $N$ critical exponents are known to be in agreement with those high order
results. More specifically the exponents $\eta$, $\eta_\phi$ and $1/\nu$ 
corresponding to the fermion and scalar field dimensions as well as the 
correlation length exponent respectively are each known to $O(1/N^2)$, 
\cite{17,24}. However, for $\eta$ this means only the first {\em two} terms in 
$1/N$ are available since the fermion has zero canonical dimension. However the
large $N$ conformal bootstrap formalism of \cite{20} allows one to determine
the $O(1/N^3)$ term of $\eta$. This has been achieved for the Ising Gross-Neveu
model in \cite{25,26,27}, as well as a variant known as the chiral Heisenberg
Gross-Neveu model in \cite{28} but not yet for the chiral XY case. This is due 
to a technical reason. In the large $N$ conformal bootstrap technique, that is 
founded on the early work of \cite{29,30,31,32}, the expansion is in powers of
the vertex anomalous dimensions. In particular in one of the amplitude 
variables the vertex anomalous dimension appears in the denominator. However in
the chiral XY Gross-Neveu case the leading order large $N$ vertex exponent is 
zero. So one cannot immediately effect the formalism at the outset. In reality 
one has a mathematically ill-defined quantity akin to $0/0$. However since the 
application of the original formalism of \cite{28,29} to the chiral XY model 
can indeed produce non-singularly derived expressions for $\eta$ to $O(1/N^2)$,
\cite{24}, there ought to be a way of applying the general large $N$ conformal 
bootstrap technology of \cite{20} as well.

Given the need for having as accurate as possible data on the phase transition
exponents for the chiral XY model we have found an indirect method to determine
$\eta$ at $O(1/N^3)$ in $d$-dimensions. This is the main topic of the article 
and it is worth outlining our strategy at the outset. Instead of considering 
the chiral XY model itself we will carry out a computation in a generalized 
theory which possesses a non-abelian symmetry. In essence this is the 
Nambu-Jona-Lasinio model, \cite{33}. Critical exponents have been computed in 
the large $N$ expansion to high order in several cases previously. For 
instance, the exponents of the fields for the group $SU(\Nc)$ have been
determined at $O(1/N^2)$ as well as that for the correlation length, 
\cite{24,34}. A calculation of $\eta$ at $O(1/N^3)$ for the $SU(2)$ 
Nambu-Jona-Lasinio model was given in \cite{35}. However we will consider the 
case of a general Lie group rather than specific unitary ones. The reason for 
this is that then the exponents will depend on the general colour Casimir group 
invariants such as the standard ones of $C_F$ and $C_A$ as well as higher rank 
ones. In this way previous results should be accessible in various limits. En 
route though we will identify several errors in earlier results. Unlike the 
present situation, when the earlier exponents were determined no high order 
$\epsilon$-expansion perturbative results were available to check against. That
is not the case now since four loop chiral XY model results are available, 
\cite{15}. Large $N$ exponents for that model are subsequently deduced from the
general non-abelian Nambu-Jona-Lasinio model by simply taking the abelian 
limit. It will be evident from the expression for $\eta$ at $O(1/N^3)$ that 
there are no singularities meaning that a smooth abelian limit can be taken. 
Indeed we will be able to show the origin of the problem in the direct large 
$N$ conformal bootstrap application to the chiral XY model. As a corollary we 
will provide an improved estimate for $\eta$ in three dimensions to the same 
number of terms as $\beta_\phi$ and $\nu$.

The article is organized as follows. We discuss the Lagrangians of the various
theories we will consider for the large $N$ analysis in Section $2$ where the 
large $N$ critical point formalism is briefly reviewed. This is applied in the
following section to determine the fermion critical exponent at $O(1/N^2)$ in
the Nambu-Jona-Lasinio model. Subsequently we complete the evaluation of both
scalar field anomalous dimensions at $O(1/N^2)$ in Section $4$ by computing 
the vertex anomalous dimensions. These and the other exponents are necessary
for the $O(1/N^2)$ evaluation of the correlation length exponent $1/\nu$ which
is carried out in Section $5$. The focus then alters to the large $N$ conformal
bootstrap formalism with the derivation of the formal underlying equations 
provided in Section $6$. The details are necessary due to the presence of more 
than one vertex unlike previous large $N$ applications of the technique. The 
actual evaluation of $\eta$ at $O(1/N^3)$ is given in Section $7$ prior to 
extracting the corresponding expression for the chiral XY model in the abelian 
limit in Section $8$. Estimates of the exponents in three dimensions are also
provided there. Finally, we provide concluding remarks in Section $9$.

\sect{Background.}

To begin with we introduce the theories we will focus on. First the four
dimensional chiral XY model has the renormalizable Lagrangian
\begin{equation}
L^{\mbox{\footnotesize{XY}}} ~=~ i \bar{\psi}^i \partialslash \psi^i ~+~
\frac{1}{2} \left( \partial_\mu \tilde{\sigma} \right)^2 ~+~
\frac{1}{2} \left( \partial_\mu \tilde{\pi} \right)^2 ~+~
g_1 \bar{\psi}^i \left( \tilde{\sigma} 
+ i \pi \gamma^5 \right) \psi^i ~+~ \frac{1}{24} g_2^2 
\left( \tilde{\sigma}^2 + \tilde{\pi}^2 \right)^2
\label{lagxyd4}
\end{equation}
where $1$~$\leq$~$i$~$\leq$~$N$ introduces our expansion parameter. The
renormalization group functions of (\ref{lagxyd4}) are available in \cite{15} 
and the $\epsilon$ expansion of the associated critical exponents at the 
Wilson-Fisher fixed point can be summed up to obtain estimates of the 
corresponding theory in three dimensions. Here our convention for $\epsilon$ is
defined by $d$~$=$~$4$~$-$~$2\epsilon$. Underlying the Wilson-Fisher fixed 
point is a universal theory which is driven by a core interaction that is 
relevant at criticality in all dimensions. It is this universal theory that one
applies the large $N$ formalism of \cite{18,19,20} to as the expansion 
parameter $1/N$ is dimensionless in all space-time dimensions. In the case of 
(\ref{lagxyd4}) the universal Lagrangian is 
\begin{equation}
L^{\mbox{\footnotesize{XY}}}_{\mbox{\footnotesize{U}}} ~=~ 
i \bar{\psi}^i \partialslash \psi^i ~+~ \sigma \bar{\psi}^i\psi^i
~+~ i \pi \bar{\psi}^i \gamma^5 \psi^i ~-~ 
\frac{1}{2g^2} \left( \sigma^2 + \pi^2 \right) ~.
\label{lagxyuniv}
\end{equation}
The scalar fields $\sigma$ and $\pi$ are related to $\tilde{\sigma}$ and
$\tilde{\pi}$ of (\ref{lagxyd4}) respectively by a coupling constant rescaling.
This is because at criticality the coupling constant is in effect fixed and
plays no role due to the scaling and conformal symmetry. In (\ref{lagxyuniv})
each field has a dimension that comprises a canonical part and an anomalous
piece. The former is determined from ensuring the action is dimensionless in
$d$-dimensions and we define the full dimensions of the three fields by
\begin{equation}
\alpha_\psi ~=~ \mu ~+~ \half \eta ~~~,~~~ 
\beta_\sigma ~=~ 1 ~-~ \eta ~-~ \chi_\sigma ~~~,~~~ 
\beta_\pi ~=~ 1 ~-~ \eta ~-~ \chi_\pi
\end{equation}
where $\eta$ is the fermion anomalous dimension with $\chi_\sigma$ and
$\chi_\pi$ corresponding to the anomalous dimensions of the respective $\sigma$
and $\pi^a$ vertex operators. At this stage we assume that the latter two are 
unequal prior to their determination. 

The problem, however, is that at leading order in $1/N$ both $\chi_\sigma$ and
$\chi_\pi$ are zero, \cite{24}. As the large $N$ conformal bootstrap programme 
is based on an expansion in the vertex anomalous dimension, for 
(\ref{lagxyuniv}) the expansion cannot begin. This is because the factor 
associated with the Polyakov conformal triangle, that is at the core of each
vertex in the large $N$ bootstrap formalism, depends on $1/\chi_\sigma$ and 
$1/\chi_\pi$ which we will demonstrate later. While previous work produced 
various large $N$ exponents to $O(1/N^2)$, \cite{24,27}, and showed that the 
vertex exponents are non-zero at $O(1/N^2)$ for (\ref{lagxyuniv}), a strategy 
needs to be devised to avoid the problem of this singularity in the bootstrap 
construction at leading order. The direction we have chosen to go in is to 
consider a more general theory which contains the universality class 
(\ref{lagxyuniv}) in a specific limit and this is the non-abelian 
Nambu-Jona-Lasinio model which has the universal Lagrangian, \cite{34}, 
\begin{equation}
L^{\mbox{\footnotesize{NJL}}}_{\mbox{\footnotesize{U}}} ~=~ 
i \bar{\psi}^{iI} \partialslash \psi^{iI} ~+~ 
\sigma \bar{\psi}^{iI} \psi^{iI} ~+~ 
i \pi^a \bar{\psi}^{iI} \gamma^5 T^a_{IJ} \psi^{iJ}
~-~ \frac{1}{2g^2} \left( {\sigma^{}}^2 + {\pi^a}^2 \right) ~.
\label{lagnjluniv}
\end{equation}
Here the fermion field has a non-abelian symmetry which we take to be any Lie 
group, classical or exceptional, with generators $T^a$ and structure constants 
$f^{abc}$ while the index ranges are $1$~$\leq$~$a$~$\leq$~$\NA$ and 
$1$~$\leq$~$I$~$\leq$~$\Nc$ where $\Nc$ and $\NA$ are the respective dimensions
of the fundamental and adjoint representations and $\pi^a$ is a vector in group
space. The four dimensional equivalent to (\ref{lagxyd4}) is
\begin{eqnarray}
L^{\mbox{\footnotesize{NJL}}} &=& i \bar{\psi}^{iI} \partialslash \psi^{iI} ~+~
\frac{1}{2} \left( \partial_\mu \tilde{\sigma} \right)^2 ~+~
\frac{1}{2} \left( \partial_\mu \tilde{\pi}^a \right)^2 \nonumber \\
&& +~ g_1 \bar{\psi}^{iI} \left( \tilde{\sigma} \delta_{IJ} 
+ i \tilde{\pi}^a \gamma^5 T^a_{IJ} \right) \psi^{iJ} ~+~ 
\frac{1}{24} g_2^2 \left( \tilde{\sigma}^2 + \tilde{\pi}^{a\,2} \right)^2
\label{lagnjld4}
\end{eqnarray}
which has a similar coupling structure to (\ref{lagxyd4}).

Comparing (\ref{lagxyuniv}) and (\ref{lagnjluniv}) the main difference is in
the decoration of fields with indices leading to Feynman rules with Lie group
dependence. Structurally in terms of Feynman graphs that will arise the only 
difference is the appearance of group factors. These will depend on the usual 
group Casimirs which are defined by
\begin{equation}
\mbox{Tr} \left( T^a T^b \right) ~=~ T_F \delta^{ab} ~~,~~
T^a T^a ~=~ C_F I_{\Nc} ~~,~~ f^{acd} f^{bcd} ~=~ C_A \delta^{ab}
\label{casdef}
\end{equation}
where $I_{\Nc}$ is the $\Nc$~$\times$~$\Nc$ unit matrix. From the equations 
defining $C_F$ and $C_A$ we have the relation
\begin{equation}
C_F \Nc ~=~ T_F \NA ~.
\end{equation}
However at high loop order it is known that new higher rank Lie group Casimirs
can appear. For instance, these first occur in a four dimensional gauge theory
in the four loop $\beta$-function of QCD, \cite{36}. For (\ref{lagnjluniv}) it 
transpires that the same feature is present in higher order large $N$ exponents
and in particular the square of the fully symmetric tensor $d_F^{abcd}$ arises 
where
\begin{equation}
d_F^{abcd} ~=~ \frac{1}{6} \mbox{Tr} \left( T^a T^{(b} T^c T^{d)} \right) ~.
\label{df4def}
\end{equation}
As will be evident from our final expressions the large $N$ exponents of 
(\ref{lagnjluniv}) will depend on the group entities $T_F$, $C_F$, $C_A$, 
$\Nc$, $\NA$ and $d_F^{abcd} d_F^{abcd}$. In light of this and observing that 
it is possible to extract (\ref{lagxyuniv}) from (\ref{lagnjluniv}) by formally
setting $T^a$~$\to$~$1$ then the critical exponents of (\ref{lagxyuniv}) should
emerge from those of (\ref{lagnjluniv}) in the abelian limit
\begin{equation}
C_F ~ \to ~ 1 ~~~,~~~
T_F ~ \to ~ 1 ~~~,~~~
\Nc ~ \to ~ 1 ~~~,~~~
C_A ~ \to ~ 0 ~~~,~~~
d_F^{abcd} d_F^{abcd} ~ \to ~ 1 ~.
\label{xylim}
\end{equation}
This then completes our computational strategy. We will focus in the main
throughout on determining the large $N$ critical exponents of 
(\ref{lagnjluniv}) before finding those of (\ref{lagxyuniv}) as a corollary via
(\ref{xylim}).

Having justified why our main focus will be on (\ref{lagnjluniv}) we recall
the basic formalism for the large $N$ critical point computations, 
\cite{18,19}. It is centred on the asymptotic behaviour of the propagators in 
the limit to the Wilson-Fisher fixed point in $d$-dimensions where $d$ does 
{\em not} play the role of a regulator. Representing the asymptotic scaling 
form of each propagator by the name of its field, the coordinate space forms 
are, \cite{37},
\begin{eqnarray}
\psi(x) &\sim& \frac{A_\psi\xslash}{(x^2)^{\alpha_\psi}} 
\left[ 1 + A_\psi^\prime(x^2)^\lambda \right] ~~,~~
\sigma(x) ~\sim~ \frac{B_\sigma}{(x^2)^{\beta_\sigma}}
\left[ 1 + B_\sigma^\prime(x^2)^\lambda \right] \nonumber \\
\pi(x) &\sim& \frac{C}{(x^2)^{\beta_\pi}} \left[ 1 + B_\pi^\prime (x^2)^\lambda
\right] 
\label{asyprop}
\end{eqnarray}
where $A_\psi$, $B_\sigma$ and $B_\pi$ are $x$-independent but $d$-dependent
amplitudes. These will appear in two combinations which we define by
\begin{equation}
z ~=~ A_\psi^2 B_\sigma ~~~,~~~ y ~=~ A_\psi^2 B_\pi
\end{equation}
and together with $\eta$ and the other exponents can be expanded in powers of
$1/N$ through
\begin{equation}
\eta(\mu) ~=~ \sum_{n=1}^\infty \frac{\eta_n(\mu)}{N^n} ~~~,~~~
z(\mu) ~=~ \sum_{n=1}^\infty \frac{z_n(\mu)}{N^n} ~~~,~~~
y(\mu) ~=~ \sum_{n=1}^\infty \frac{y_n(\mu)}{N^n} 
\label{paramexp}
\end{equation}
for example. In addition to the leading terms of (\ref{asyprop}) we have
included corrections to scaling corresponding to the terms with
$(x^2)^\lambda$ and associated amplitudes $A_\psi^\prime$, $B_\sigma^\prime$ 
and $B_\pi^\prime$. The exponent $\lambda$ is usually used to determine the
exponent $1/\nu$ which is related to the correlation length and we will 
consider it here too. In that case the canonical value of $\lambda$ is 
$(\mu-1)$. In order to find the first few orders of each exponent requires the 
solution of skeleton Schwinger-Dyson $2$-point functions which require the
asymptotic scaling counterparts to the propagators of (\ref{asyprop}). They 
were derived in \cite{37} by inverting their momentum space forms using the 
Fourier transform which is given by, 
\cite{18}, 
\begin{equation}
\frac{1}{(x^2)^\alpha} ~=~ \frac{a(\alpha)}{2^{2\alpha}} \int_k
\frac{e^{ikx}}{(k^2)^{\mu-\alpha}}
\label{fourier}
\end{equation}
in general to set notation where
\begin{equation}
a(\alpha) ~=~ \frac{\Gamma(\mu-\alpha)}{\Gamma(\alpha)} ~.
\end{equation}
Consequently we have 
\begin{eqnarray}
\psi^{-1}(x) & \sim & 
\frac{r(\alpha_\psi-1)\xslash}{A_\psi(x^2)^{2\mu-\alpha_\psi+1}}
\left[ 1 - A_\psi^\prime s(\alpha_\psi-1)(x^2)^\lambda \right] \nonumber \\
\sigma^{-1}(x) & \sim &\frac{p(\beta_\sigma)}
{B_\sigma (x^2)^{2\mu-\beta_\sigma}}
\left[ 1 - B_\sigma^\prime q(\beta_\sigma) (x^2)^\lambda \right] \nonumber \\
\pi^{-1}(x) & \sim &\frac{p(\beta_\pi)}
{B_\pi (x^2)^{2\mu-\beta_\pi}}
\left[ 1 - B_\pi^\prime q(\beta_\pi) (x^2)^\lambda \right]
\end{eqnarray}
where the various functions are defined by 
\begin{eqnarray}
p(\beta) &=& \frac{a(\beta-\mu)}{a(\beta)} ~~~~~~,~~~~~~
r(\alpha) ~=~ \frac{\alpha p(\alpha)}{(\mu-\alpha)} \nonumber \\
q(\beta) &=& \frac{a(\beta-\mu+\lambda)a(\beta-\lambda)}
{a(\beta-\mu)a(\beta)} ~~~,~~~ 
s(\alpha) ~=~ \frac{\alpha(\alpha-\mu)q(\alpha)}{(\alpha-\mu+\lambda)
(\alpha-\lambda)} 
\label{def2ptscal}
\end{eqnarray}
for arbitrary $\alpha$, $\beta$ and $\lambda$.

\sect{Evaluation of $\eta_2$.}

The asymptotic scaling forms of the propagators are necessary in order to
algebraically represent the behaviour of the Schwinger-Dyson equations in the 
critical region. To determine $\eta$ at $O(1/N^2)$ we consider the $2$-point 
functions of the three fields of (\ref{lagxyuniv}) and to the order we are 
interested the relevant graphs that contribute are provided in Figures 
\ref{figpsi}, \ref{figsig} and \ref{figpi}. In each there are no self-energy
corrections on any of the propagators. This is because the propagator powers
include the non-zero anomalous dimensions $\eta$, $\chi_\sigma$ and $\chi_\pi$
that account for such effects. In terms of counting with respect to the 
ordering parameter $1/N$, each closed fermion loop contributes a power of $N$ 
whereas a $\sigma$ or $\pi^a$ field has a factor of $1/N$ associated with it. 
This is accounted for through the $N$ dependence of the amplitude combination 
variables $z$ and $y$ defined in (\ref{paramexp}). So, for example, all the two
loop graphs differ from the respective one loop graphs of their Schwinger-Dyson
equations by a power of $1/N$. For the scalar $O(N)$ theory of \cite{18,19} 
various three loop graphs also contribute at $O(1/N^2)$ but the corresponding 
graphs are absent here. This is because the extra diagrams have closed fermion 
loops with three scalars external to that loop. Such graphs are zero because of
the trace over an odd number of $\gamma$-matrices. Substituting the propagators
into the skeleton Schwinger-Dyson equation for $\psi$ equates algebraically to 
\begin{eqnarray}
0 &=& r(\alpha-1) ~+~ 
z Z_{V_\sigma}^2 (x^2)^{\chi_\sigma+\Delta}  ~-~
C_F y Z_{V_\pi}^2 (x^2)^{\chi_\pi+\Delta} \nonumber \\
&& +~ \left[
z^2 (x^2)^{2\chi_\sigma+2\Delta} ~-~
2 C_F z y (x^2)^{\chi_\sigma+\chi_\pi+2\Delta} ~+~
C_F ( C_F - \half C_A ) y^2 (x^2)^{2\chi_\pi+2\Delta}
\right] \Sigma_1 \nonumber \\
&& +~ O \left( \frac{1}{N^3} \right) 
\label{sdepsi}
\end{eqnarray}
where we have included the various group theory factors associated with the
non-abelian symmetry, as well as the vertex renormalization constants 
$Z_{V_\sigma}$ and $Z_{V_\pi}$, and effected the $\gamma$-algebra. These will 
have poles in the analytic regularization $\Delta$ that is introduced through 
the replacement, \cite{18,19},
\begin{equation}
\chi_\sigma ~\rightarrow~ \chi_\sigma ~+~ \Delta ~~~,~~~ 
\chi_\pi ~\rightarrow~ \chi_\pi ~+~ \Delta ~.
\end{equation}
Such a regulator is required since the actual two loop corrections themselves
represented by $\Sigma_1$ for the $\psi$ equation are divergent.

{\begin{figure}[hb]
\begin{center}
\includegraphics[width=10cm,height=6.0cm]{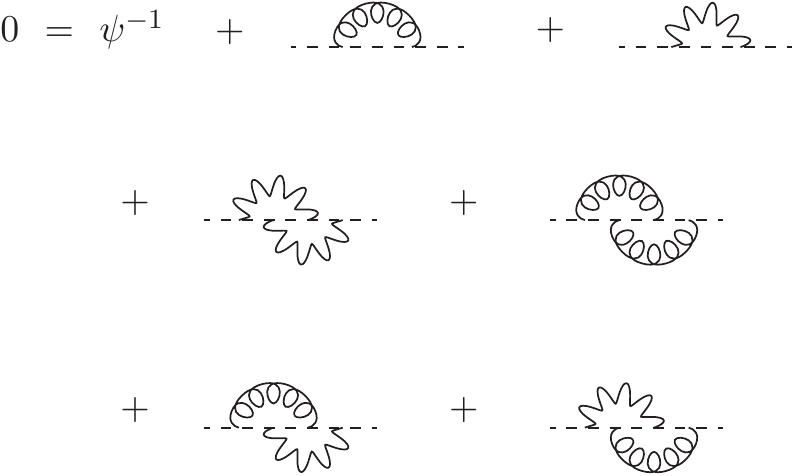}
\end{center}
\caption{Skeleton Schwinger-Dyson $2$-point function for $\psi$ at 
$O(1/N^2)$.}
\label{figpsi}
\end{figure}}

Once the group and amplitude dependence are factored off the remaining 
contribution is the Feynman integral itself. Its value is the same irrespective
of which of the two interactions are involved. Expressions for both $\Sigma_1$ 
and $\Pi_1$ are provided in \cite{37} where the latter arises in Figures 
\ref{figsig} and \ref{figpi}. The representation of the respective skeleton 
Schwinger-Dyson equations analogous to (\ref{sdepsi}) are
\begin{eqnarray}
0 &=& p(\beta_\sigma) ~+~ N \Nc z Z_{V_\sigma}^2 (x^2)^{\chi_\sigma+\Delta}
\nonumber \\
&& -~ \frac{N}{2} \left[ \Nc z^2 (x^2)^{2\chi_\sigma+2\Delta} ~-~
C_F \Nc y z (x^2)^{\chi_\sigma+\chi_\pi+2\Delta} 
\right] \Pi_1 +~ O \left( \frac{1}{N^2} \right)
\label{sdesig}
\end{eqnarray}
and
\begin{eqnarray}
0 &=& p(\beta_\pi) ~-~ N T_F y Z_{V_\pi}^2 (x^2)^{\chi_\pi+\Delta}
\nonumber \\
&& -~ \frac{N}{2} T_F \left[ y^2 (x^2)^{2\chi_\pi+2\Delta} ~-~
( C_F - \half C_A ) y z (x^2)^{\chi_\sigma+\chi_\pi+2\Delta} 
\right] \Pi_1 +~ O \left( \frac{1}{N^2} \right) ~. 
\label{sdepi}
\end{eqnarray}
In all three cases we have not included vertex renormalization constants in the
higher order terms since those counterterms would contribute to the next order 
in the respective expansions. Also we have cancelled off a common factor of 
$x^2$ whose power is related to the canonical dimension. What remains are 
factors of $x^2$ whose exponent involves a linear combination of the regulator 
and the vertex anomalous dimensions. These cannot be neglected despite being 
small. This is because in the approach to criticality there is no overall 
scale. For instance as $x^2$~$\to$~$0$
\begin{equation}
(x^2)^{\chi_\sigma+\Delta} ~=~ 1 ~+~ \frac{\chi_\sigma}{N} \ln(x^2) ~+~
\Delta \ln(x^2) ~+~ O \left( \Delta; \frac{1}{N^2} \right)
\end{equation}
to the order of interest. If the $\ln(x^2)$ terms remain then they would give
singularities in the limit as $x^2$~$\to$~$0$ or $\infty$. However the simple
pole in each of $\Sigma_1$ and $\Pi_1$ is removed by the vertex counterterm. 
Once this is fixed the remaining terms involve $\chi_{\sigma\,1}$ or 
$\chi_{\pi\,1}$ as well as a term dependent on the counterterm but where each 
is multiplied by $\ln(x^2)$. These terms are removed by defining the respective
vertex anomalous dimensions. A consistency check on this is that the same 
solution for each of $\chi_{\sigma\,1}$ and $\chi_{\pi\,1}$ emerges from all 
three equations.

{\begin{figure}[ht]
\begin{center}
\includegraphics[width=9.25cm,height=3.5cm]{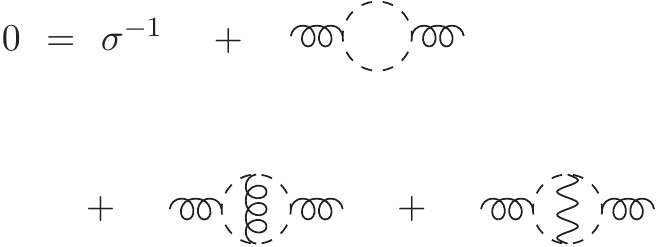}
\end{center}
\caption{Skeleton Schwinger-Dyson $2$-point function for $\sigma$ at 
$O(1/N^2)$.}
\label{figsig}
\end{figure}}

{\begin{figure}[hb]
\begin{center}
\includegraphics[width=9.25cm,height=3.5cm]{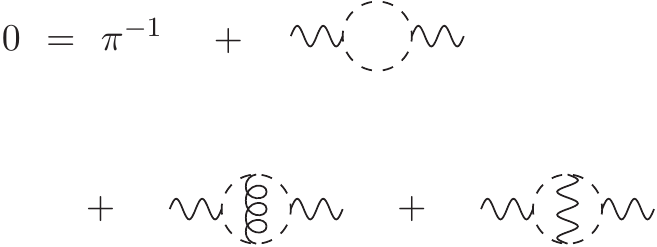}
\end{center}
\caption{Skeleton Schwinger-Dyson $2$-point function for $\pi$ at $O(1/N^2)$.}
\label{figpi}
\end{figure}}

Eliminating $z$ and $y$ at successive orders in $1/N$ from (\ref{sdepsi}),
(\ref{sdesig}) and (\ref{sdepi}) and expanding the scaling functions for the 
$2$-point functions of (\ref{def2ptscal}), we find 
\begin{equation}
\eta_1 ~=~ -~ \frac{2 [T_F+C_F\Nc] \Gamma(2\mu-1)}
{\mu \Gamma(\mu-1) \Gamma(1-\mu) \Gamma^2(\mu) \Nc T_F}
\label{njleta1}
\end{equation}
at leading order. From the $O(1/N^2)$ parts of the equation, after 
renormalization we deduce
\begin{equation}
\chi_{\sigma\,1} ~=~ -~ \frac{\mu [ C_F \Nc - T_F ] \eta_1}
{[T_F+C_F \Nc][\mu-1]} ~~,~~
\chi_{\pi\,1} ~=~ -~ \frac{\mu [ 2 T_F - 2 C_F \Nc + C_A \Nc ] \eta_1}
{2[\mu-1] [T_F+C_F \Nc]}
\label{njlchi1}
\end{equation}
to ensure no $\ln(x^2)$ terms remain resulting in
\begin{eqnarray}
\eta_2 &=&
\left[
[ 2 (2 \mu-1) ( C_F^2 \Nc^2 + T_F^2 )
- 4 C_F \Nc T_F - \mu C_F C_A \Nc^2 ]
\frac{\Psi(\mu)}{2[\mu-1]}
\right. \nonumber \\
&& \left. ~
+ \frac{(2 \mu-1) (4 \mu-1)}{2\mu[\mu-1]^2} [ C_F^2 \Nc^2 + T_F^2 ]
+ \frac{( 2 \mu^2 - 6 \mu + 1 )}{\mu[\mu-1]^2} C_F \Nc T_F
\right. \nonumber \\
&& \left. ~
- \frac{3 \mu}{4[\mu-1]^2} C_F C_A \Nc^2 
\right]
\frac{\eta_1^2}{[T_F+C_F \Nc]^2}
\label{njleta2}
\end{eqnarray}
from the finite part of the consistency equations. Here we use the shorthand
notation
\begin{equation}
\Psi(\mu) ~=~ \psi(2\mu-1) - \psi(1) + \psi(2-\mu) - \psi(\mu)
\end{equation}
where $\psi(z)$ is the Euler $\psi$ function. 

\sect{Vertex anomalous dimensions at $O(1/N^2)$.}

One of the key ingredients to proceeding to higher order in the large $N$ 
expansion is the provision of the vertex anomalous dimensions. The general 
approach is similar to conventional perturbation theory in that at successive 
loop orders the wave function renormalization constants are needed first prior 
to renormalizing the mass and coupling constants. While the fermion anomalous 
dimension is associated with the critical exponent $\eta$ those for the scalar 
fields require $\chi_\sigma$ and $\chi_\pi$ and we summarize their
determination in this section. While $\chi_{\sigma\,1}$ and $\chi_{\pi\,1}$ 
were deduced as a corollary to finding $\eta_2$ by excluding $\ln(x^2)$ terms
thereby ensuring the scaling limit could be taken smoothly, the vertex 
dimensions can also be computed directly from the two vertex $3$-point 
functions. The leading order $1/N$ corrections for the $\sigma \bar{\psi}\psi$ 
vertex are shown in Figure \ref{figchi1}. Those for the $\pi^a \bar{\psi} 
\gamma^5 T^a \psi$ are virtually the same but with the modification that the 
external $\sigma$ line is replaced by one representing $\pi^a$. It is for this 
reason that we have not illustrated them.  

{\begin{figure}[ht]
\begin{center}
\includegraphics[width=5.5cm,height=2.0cm]{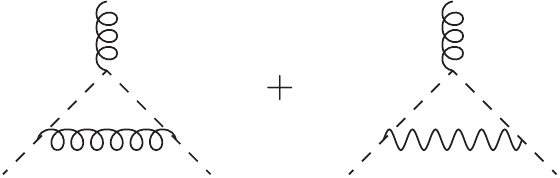}
\end{center}
\caption{Leading order corrections to $\sigma \bar{\psi}\psi$ $3$-point 
function.}
\label{figchi1}
\end{figure}}

To extract expressions for $\chi_{\sigma\,1}$ and $\chi_{\pi\,1}$ from Figure 
\ref{figchi1} we follow the method of \cite{18,19,38,39} which is the large $N$
critical point renormalization formalism. One uses the asymptotic scaling forms
of the propagators (\ref{asyprop}) but $\Delta$ regularized. This produces an 
expansion in powers of $1/N$ where at each order the terms are a Laurent series
in $\Delta$. The poles corresponding to the divergences are absorbed into the 
vertex renormalization constants $Z_{V_\sigma}$ and $Z_{V_\pi}$ in general 
although the $O(1/N)$ terms of each are already available. Once the divergences
are removed the remaining finite part does not have a non-singular limit to 
criticality. By this we mean that in coordinate space $\ln(x^2)$ parts are 
present. Equally carrying out the computation in momentum space the partner 
$\ln(p^2)$ dependence remains. In each case there are terms which involve 
$\chi_{\sigma\,1} \ln(x^2)$ and $\chi_{\pi\,1} \ln(x^2)$. Thus to ensure a 
scale free finite Green's function these unknowns are chosen to remove the 
residual $x^2$ or $p^2$ dependence. In the way we have presented the larger 
formalism in the previous section, the values that we extract for 
$\chi_{\sigma\,1}$ and $\chi_{\pi\,1}$ from the graphs of Figure \ref{figchi1} 
fully agree with (\ref{njlchi1}).

{\begin{figure}[ht]
\begin{center}
\includegraphics[width=14.0cm,height=9.5cm]{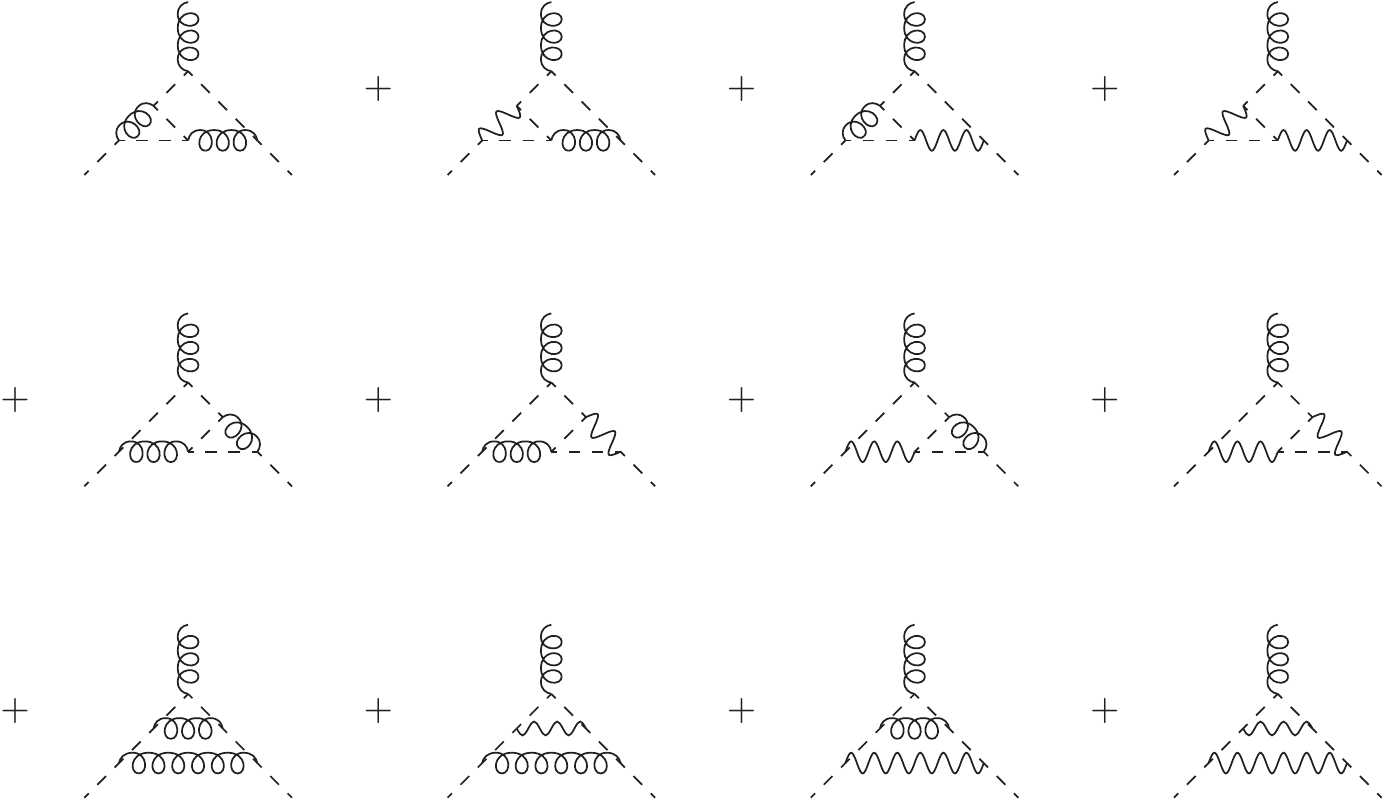}
\end{center}
\caption{Vertex corrections to $\sigma \bar{\psi}\psi$ $3$-point function at 
$O(1/N^2)$.}
\label{figchi2a}
\end{figure}}

At $O(1/N^2)$ there are a considerably larger number of diagrams to evaluate.
For instance, each of the three vertices in both graphs of Figure \ref{figchi1}
will gain vertex corrections. These are displayed in Figure \ref{figchi2a}
where again we will only illustrate the graphs for the $\sigma \bar{\psi}\psi$
$3$-point function. Those for the $\pi^a \bar{\psi} \gamma^5 T^a \psi$ are 
obtained by swapping the external scalar in each graph. In addition to these 
there will be contributions from expanding the graphs of Figure \ref{figchi1} 
out to $O(1/N^2)$ due to the $N$ dependence in the anomalous part of the 
propagator exponents in each graph. More significantly the set of graphs in 
Figure \ref{figchi2b} need to be included. These are shown separately, partly 
for a reason that will become evident later, but mainly because they can be 
regarded as primitive in the sense that the highest order of pole in $\Delta$ 
is simple. One aspect of the evaluation of the graphs of Figure \ref{figchi2b} 
requiring care is that of determining the group theory factor associated with 
the three loop light-by-light graphs. This is particularly more involved for 
the $\pi^a \bar{\psi} \gamma^5 T^a \psi$ vertex function when all the scalar 
fields are $\pi^a$ ones. In this instance one has to rationalize traces of the 
form $\mbox{Tr} \left( T^a T^b T^c T^d \right)$ for a general Lie group. To 
handle this we have used the {\tt color.h} module that is written in the 
symbolic manipulation language {\sc Form}, \cite{40,41}. Indeed we have used 
{\sc Form} to handle the algebra associated with solving all the various 
Schwinger-Dyson self-consistency equations. The {\tt color.h} routine is an 
encoding of the Lie group Casimir analysis of \cite{42} that goes beyond the 
rank $2$ ones of (\ref{casdef}). In particular the earlier trace can be 
rewritten in terms of $d_F^{abcd} d_F^{abcd}$ and $d^{abc} d^{abc}$ where 
\begin{equation}
d^{abc} ~=~ \frac{1}{2} \mbox{Tr} \left( T^a T^{(b} T^{c)} \right)
\end{equation}
is fully symmetric. Though for all the light-by-light graphs of Figure 
\ref{figchi2b} the combination $d^{abc} d^{abc}$ cancels in the final
expressions for not only $\chi_{\sigma\,2}$ and $\chi_{\pi\,2}$ but also all
the $O(1/N^2)$ and higher order exponents where $d_F^{abcd} d_F^{abcd}$ occurs. 

{\begin{figure}[ht]
\begin{center}
\includegraphics[width=14.0cm,height=9.5cm]{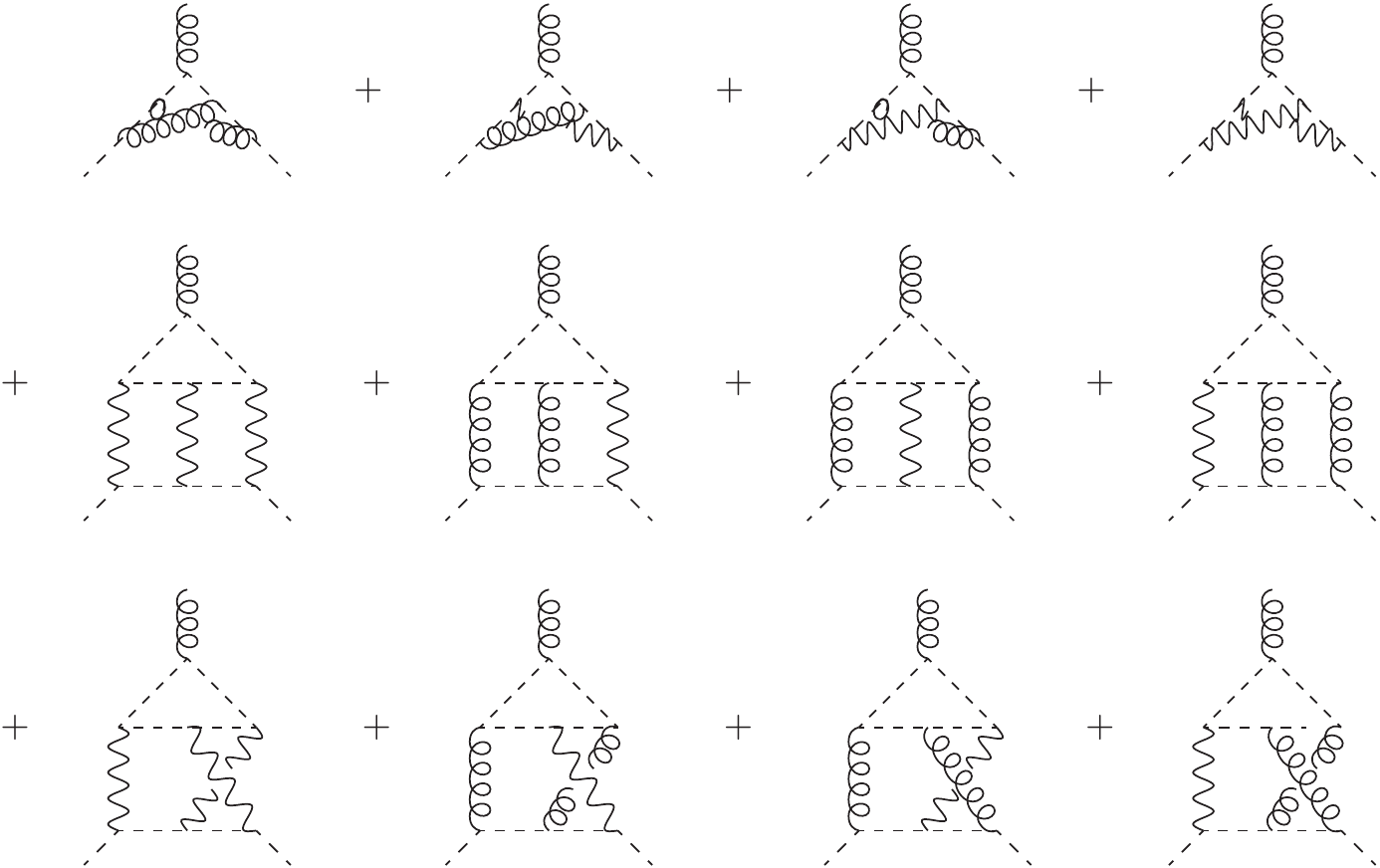}
\end{center}
\caption{Additional corrections to $\sigma \bar{\psi}\psi$ $3$-point function 
at $O(1/N^2)$.}
\label{figchi2b}
\end{figure}}

Assembling the values of the various diagrams with their associated group 
theory factor and extracting the piece associated with the separate vertex 
exponents we find 
\begin{eqnarray}
\chi_{\sigma\,2} &=& \left[
\frac{3 \mu^2}{[\mu-1]} [ T_F^2 - C_F \Nc T_F ] \Theta(\mu)
- \frac{\mu}{[\mu-1]^2} C_F^2 \Nc^2
- \frac{\mu^2 (6 \mu^2-17 \mu+14)}{[\mu-1]^3} C_F \Nc T_F
\right. \nonumber \\
&& \left. ~
- \frac{\mu (2 \mu-1) (\mu^2-\mu-1)}{[\mu-1]^3} T_F^2
- \frac{\mu\Psi(\mu)}{2[\mu-1]^2}
\left[ 2 (2 \mu-1) [ C_F^2 \Nc^2 - T_F^2 ] - \mu C_F C_A \Nc^2 \right]
\right] \nonumber \\
&& \times \frac{\eta_1^2}{[T_F+C_F \Nc]^2}
\label{chisig2}
\end{eqnarray}
from the graphs of Figure \ref{figchi2a} and \ref{figchi2b} where
\begin{equation}
\Theta(\mu) ~=~ \psi^\prime(\mu) ~-~ \psi^\prime(1) ~. 
\end{equation}
Similarly we find
\begin{eqnarray}
\chi_{\pi \, 2} &=&
\left[ \frac{\mu\Psi(\mu)}{4[\mu-1]^2}
\left[ 4 (2 \mu-1) ( C_F^2 \Nc^2 - T_F^2 )
- 2 ( (3 \mu-1) C_F \Nc - T_F ) C_A \Nc
+ \mu C_A^2 \Nc^2 \right]
\right. \nonumber \\
&& \left. ~
+ \frac{\mu^2 \Theta(\mu)}{8[\mu-1]}
\left[ 24 \frac{d_F^{abcd} d_F^{abcd} \Nc}{C_F T_F}
- 24 T_F^2 - C_A^2 \Nc^2 \right]
+ \frac{\mu (2 \mu-1)}{[\mu-1]^3} C_F^2 \Nc^2
+ \frac{3 \mu^2}{[\mu-1]^3} C_F \Nc T_F
\right. \nonumber \\
&& \left. ~
- \frac{\mu^2 (2 \mu-1)}{[\mu-1]^2}
\left[ \frac{d_F^{abcd} d_F^{abcd}\Nc}{C_F T_F} - T_F^2 \right]
- \frac{\mu (7 \mu-2)}{4[\mu-1]^3} C_F C_A \Nc^2
- \frac{\mu^2 ( 2 \mu^2 - 6 \mu + 1 )}{12[\mu-1]^3} C_A^2 \Nc^2
\right. \nonumber \\
&& \left. ~
- \frac{\mu (8 \mu^3-20 \mu^2+19 \mu-1)}{[\mu-1]^3} T_F^2
- \frac{\mu (5 \mu-2)}{4[\mu-1]^3} C_A \Nc T_F
\right] \frac{\eta_1^2}{[T_F+C_F \Nc]^2} ~.
\label{chipi2}
\end{eqnarray}
Both (\ref{chisig2}) and (\ref{chipi2}) are more involved than the expression
for $\eta_2$ primarily due to the light-by-light contributions with the term
involving $\Theta(\mu)$ arising from the non-planar ones. We note though that
similar to \cite{24} the vertex anomalous dimensions are equal at $O(1/N^2)$ 
for the $SU(2)$ group.

{\begin{figure}[ht]
\begin{center}
\includegraphics[width=11cm,height=11cm]{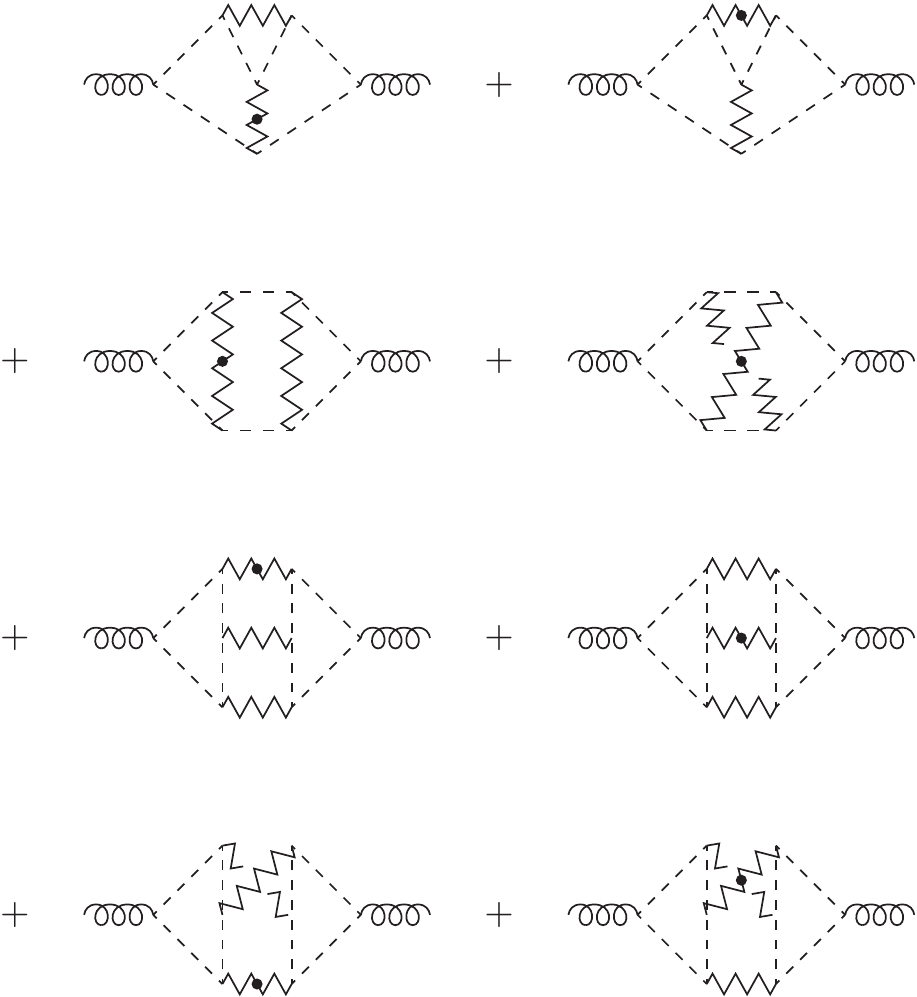}
\end{center}
\caption{Graphs for $O(1/N^2)$ correction to the $\sigma$ skeleton
Schwinger-Dyson $2$-point function to determine $\lambda_2$ using shorthand 
representation for the internal scalar fields.}
\label{figlam}
\end{figure}}

\sect{Computation of $\lambda_2$.}

Having (\ref{chisig2}) and (\ref{chipi2}) now means the scalar wave function 
anomalous dimensions are known at $O(1/N^2)$. This was achieved by using the 
leading terms of the asymptotic scaling forms of the propagators 
(\ref{asyprop}). To deduce other exponents one considers the correction to 
scaling terms and to find $\nu$ in particular at $O(1/N^2)$ we take 
$\lambda$~$=$~$1/(2\nu)$ where the canonical value of $\lambda$ is $(\mu-1)$. 
For (\ref{lagnjluniv}) we therefore include the correction term on the 
propagators in the graphs of Figures \ref{figpsi}, \ref{figsig} and \ref{figpi}
retaining only those contributions with one insertion only, \cite{19}. Unlike 
the scalar $O(N)$ model of \cite{18,19} for the Gross-Neveu universality class 
there is a reordering of the diagrams with corrections on the $\sigma$ and 
$\pi^a$ lines in both their self-consistency equations, \cite{25,26}. This 
stems from the correction term of the $2$-point functions of (\ref{def2ptscal})
and in particular $q(\beta_\sigma)$ and $q(\beta_\pi)$. Specifically the 
leading order large $N$ term of $a(\beta-\mu+\lambda)$ of (\ref{def2ptscal}) is
$O(N)$ for both $\beta_\sigma$ and $\beta_\pi$ rather than $O(1)$ for the
analogous function in the scalar theory of \cite{19}. Consequently to find 
$\nu$ at $O(1/N)$ the two graphs of Figures \ref{figsig} and \ref{figpi} with 
corrections on the $\sigma$ and $\pi^a$ internal propagators have to be 
included.

This is evident when the self-consistency equation for $\lambda$ is 
constructed. Clearly the correction to scaling terms in (\ref{asyprop}) are of 
different length dimension to the leading terms. So the leading and correction 
terms with regard to $x^2$ in the algebraic representation of the 
Schwinger-Dyson equations decouple. The leading term that allowed us to deduce 
$\eta_2$ has been dealt with but three equations are left each involving the 
correction amplitudes $A_\psi^\prime$, $B_\sigma^\prime$ and $B_\pi^\prime$. 
The three equations are combined into one equation involving a $3$~$\times$~$3$
matrix containing $\lambda_1$ as the only unknown at leading order. However 
examining the $N$ dependence of the terms in the matrix it is evident that it 
is different in different rows. Omitting the higher order graphs for the 
$\sigma$ and $\pi^a$ Schwinger-Dyson equations would have left an inconsistent 
set of equations. The final step to find $\lambda_1$ is to set the determinant 
of this matrix to zero whence we find
\begin{equation}
\lambda_1 ~=~ -~ (2\mu - 1) \eta_1 ~.
\label{njllam1}
\end{equation}

The consequence of the reordering due to the singular nature of 
$q(\beta_\sigma)$ and $q(\beta_\pi)$ is that at next order the higher order 
graphs of Figure \ref{figlam} have to be included where the dot on a scalar 
line indicates the propagator with the correction to scaling term of 
(\ref{asyprop}). Here we have introduced a shorthand notation to compress the 
number of graphs that actually contribute. In Figure \ref{figlam} the zigzag 
line represents both $\sigma$ and $\pi^a$ fields. So, for instance, this means 
that in reality there are four graphs from each possible choice zigzag line in 
the three loop graphs and eight for the four loop ones. In total that is $48$ 
graphs for each scalar field Schwinger-Dyson equation. The values of each graph
independent of the group theory factor are available in \cite{43}. However 
appending the group values the solution to the determinant of the 
self-consistency equation to next order in $1/N$ gives
\begin{eqnarray}
\lambda_2 &=& \left[ \left[ 8 (2 \mu - 1) (2 \mu - 3) C_F T_F
+ 4 \frac{d_F^{abcd} d_F^{abcd} \Nc}{T_F^2}
+ 4 \frac{T_F^2}{\Nc}
\right. \right. \nonumber \\
&& \left. \left. ~~~
+ (3 \mu^2 - 6 \mu + 2) \frac{C_F C_A^2 \Nc^2}{6T_F} \right]
\frac{\mu[T_F + C_F \Nc]}{[\mu - 1][\mu - 2]^2 \eta_1}
\right. \nonumber \\
&& \left. ~
- \left[ 4 C_F \Nc T_F^2
+ 2 T_F^3
+ \frac{1}{24} C_A^2 C_F \Nc^3
+ 2 \frac{d_F^{abcd} d_F^{abcd} \Nc^2}{T_F} \right]
\frac{\mu^2 (2 \mu - 3)}{[\mu - 1][\mu - 2]} 
\left[ \Phi(\mu) + \Psi^2(\mu) \right]
\right. \nonumber \\
&& \left. ~
+ \left[
\mu (\mu - 1) (2 \mu - 1) (\mu - 2)^2 C_F C_A \Nc^2 T_F
- (\mu + 1) (\mu - 2)^2 (\mu - 1) (2 \mu - 1)^2 C_F^3 \Nc^3
\right. \right. \nonumber \\
&& \left. \left. ~~~~~
- (2 \mu^2 - 5 \mu + 1) (2 \mu^4 - 5 \mu^3 + 4 \mu^2 - 4) T_F^3
\right. \right. \nonumber \\
&& \left. \left. ~~~~~
- (\mu - 1) (\mu - 2)^2 (2 \mu - 1) (6 \mu^2 - 5 \mu - 3) C_F^2 \Nc^2 T_F
\right. \right. \nonumber \\
&& \left. \left. ~~~~~
- (12 \mu^6 - 92 \mu^5 + 251 \mu^4 - 294 \mu^3+ 111 \mu^2 + 28 \mu - 12) 
C_F \Nc T_F^2
\right. \right. \nonumber \\
&& \left. \left. ~~~~~
+ \frac{1}{24} \mu^2 (\mu - 1) (6 \mu^2 - 21 \mu + 20) C_A^2 C_F \Nc^3
+ \mu (\mu - 1) (\mu - 2)^2 (2 \mu - 1) C_A C_F^2 \Nc^3
\right. \right. \nonumber \\
&& \left. \left. ~~~~~
- \mu^2 (3 \mu - 5) (2 \mu - 5) \frac{d_F^{abcd} d_F^{abcd} \Nc^2}{T_F} \right]
\frac{\Psi(\mu)}{[\mu - 1]^2[\mu - 2]^2}
\right. \nonumber \\
&& \left. ~
+ \left[
\frac{3}{4} \mu^2 (2 \mu - 3) (\mu - 2) C_F C_A \Nc^2 T_F
- \frac{3}{2} \mu^2 (2 \mu + 1) (\mu - 2) C_F^3 \Nc^3
\right. \right. \nonumber \\
&& \left. \left. ~~~~~
- \frac{3}{2} \mu^2 (2 \mu^2 - 13 \mu + 12) T_F^3
+ \frac{3}{2} \mu^2 (2 \mu - 3) (\mu - 2) C_F^2 \Nc^2 T_F
\right. \right. \nonumber \\
&& \left. \left. ~~~~~
+ \frac{3}{2} \mu^2 (2 \mu^2 + 5 \mu - 14) C_F \Nc T_F^2
- \frac{11}{8} \mu^2 (\mu - 2) C_A^2 C_F \Nc^3
\right. \right. \nonumber \\
&& \left. \left. ~~~~~
+ \frac{3}{4} \mu^2 (2 \mu + 5) (\mu - 2) C_A C_F^2 \Nc^3
+ 3 \mu^2 (5 \mu - 7) \frac{d_F^{abcd} d_F^{abcd} \Nc^2}{T_F} \right]
\frac{\Theta(\mu)}{[\mu - 1][\mu - 2]}
\right. \nonumber \\
&& \left. ~
+ \frac{(2 \mu - 1)^2 (2 \mu^3 - 4 \mu^2 - 2 \mu + 1)}{2\mu[\mu - 1]^2}
C_F^3 \Nc^3
+ \frac{\mu (3 \mu - 1) (2 \mu - 3)}{4[\mu - 1]^3} C_F C_A \Nc^2 T_F
\right. \nonumber \\
&& \left. ~
+ \frac{(8 \mu^8 - 68 \mu^7 + 212 \mu^6 - 292 \mu^5 + 145 \mu^4 + 48 \mu^3
- 85 \mu^2 + 32 \mu - 4)}{2\mu[\mu - 1]^3[\mu - 2]^2} T_F^3
\right. \nonumber \\
&& \left. ~
+ \frac{(24 \mu^6 - 96 \mu^5 + 126 \mu^4 - 38 \mu^3 - 30 \mu^2 + 21 \mu - 3)}
{2\mu[\mu - 1]^3} C_F^2 \Nc^2 T_F
\right. \nonumber \\
&& \left. ~
+ \frac{[24 \mu^8 - 216 \mu^7 + 730 \mu^6 - 1178 \mu^5 + 868 \mu^4 - 101 \mu^3
- 207 \mu^2 + 96 \mu - 12]}{2\mu[\mu - 1]^3[\mu - 2]^2} C_F \Nc T_F^2
\right. \nonumber \\
&& \left. ~
- \frac{\mu^2 (8 \mu^4 - 42 \mu^3 + 85 \mu^2 - 75 \mu + 20)}
{48[\mu - 1]^3[\mu - 2]^2} C_A^2 C_F \Nc^3
+ \frac{3\mu (2 \mu - 1)}{4[\mu - 1]^2} C_A C_F^2 \Nc^3
\right. \nonumber \\
&& \left. ~
- \frac{\mu^2 (4 \mu^4 - 18 \mu^3 + 26 \mu^2 - 15 \mu + 7)}
{2[\mu - 1]^3[\mu - 2]^2} \frac{d_F^{abcd} d_F^{abcd} \Nc^2}{T_F}
\right] \frac{\eta_1^2}{[T_F + C_F \Nc]^3} ~.
\label{njllam2}
\end{eqnarray}
The contributions from the light-by-light graphs of Figure \ref{figlam} are
evident.

Finally we can note that the expressions for the exponents to this stage are in
agreement with previously determined exponents in the Gross-Neveu model,
\cite{25,26,37,43}, in the limit
\begin{equation}
C_F ~ \to ~ 0 ~~~,~~~
T_F ~ \to ~ 1 ~~~,~~~
\Nc ~ \to ~ 0 ~~~,~~~
C_A ~ \to ~ 0 ~~~,~~~
d_F^{abcd} d_F^{abcd} ~ \to ~ 0
\label{gnlim}
\end{equation}
since this excludes graphs involving the $\pi^a \bar{\psi} \gamma^5 T^a \psi$ 
vertex. For the Gross-Neveu XY model taking the limit of (\ref{xylim}) we also 
find agreement with \cite{17,24}. These provide important checks on our 
exponents for (\ref{lagnjluniv}). 

{\begin{figure}[ht]
\begin{center}
\includegraphics[width=9cm,height=2.7cm]{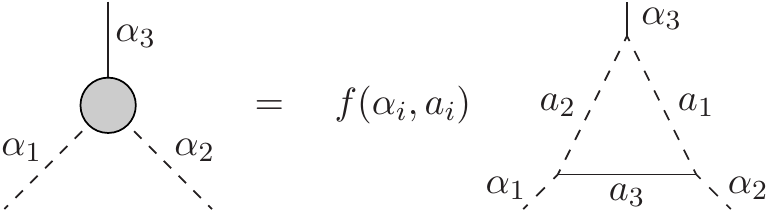}
\end{center}
\caption{Definition of internal indices in the Polyakov conformal triangle of
a scalar Yukawa interaction.}
\label{figcntr}
\end{figure}}

{\begin{figure}[hb]
\begin{center}
\includegraphics[width=12.0cm,height=1.0cm]{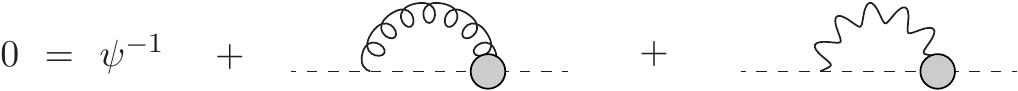}
\end{center}
\caption{Schwinger-Dyson equation for $\psi$.}
\label{figpsicb}
\end{figure}}

\sect{Large $N$ conformal bootstrap equations.}

{\begin{figure}[ht]
\begin{center}
\includegraphics[width=7.0cm,height=4.0cm]{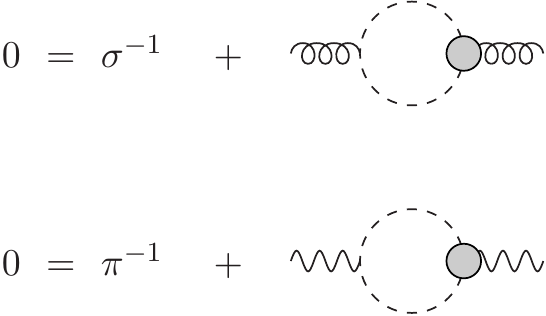}
\end{center}
\caption{Schwinger-Dyson equations for $\sigma$ and $\pi^a$.}
\label{figsigcb}
\end{figure}}

Before concentrating on the $O(1/N^3)$ evaluation of $\eta$ using the conformal
bootstrap we devote this section to recalling one of the key features of the 
formalism. This is the Polyakov conformal triangle, \cite{29,44}, which 
represents the full vertex function in the theory at criticality. Ordinarily a 
vertex operator has a non-zero anomalous dimension. In other words in 
coordinate space the sum of the exponents of the propagators joining to a 
$3$-point vertex has a non-zero part that means the overall dimension does not 
equal the canonical dimension. This is in fact the reason why the conformal 
integration rule referred to as uniqueness, \cite{19,31}, cannot directly be 
applied to the two loop graphs of Figures \ref{figpsi}, \ref{figsig} and 
\ref{figpi}. The canonical dimension of each vertex in the theory matches the 
value of $(2\mu+1)$ satisfied by the uniqueness rule of \cite{37} but 
$\chi_\sigma$ and $\chi_\pi$ have small but non-zero values. The Polyakov 
conformal triangle construction allows one to exploit uniqueness as a 
computational tool by replacing the full vertex with a one loop graph where the
exponents of the internal lines are arbitrary but are chosen to make the new 
internal vertices unique. For a Yukawa vertex the most general conformal 
triangle is shown in Figure \ref{figcntr}, \cite{20,25,26}. Specifically 
\begin{eqnarray}
a_1 ~+~ a_2 ~+~ \alpha_3 &=& 2\mu ~+~ 1 \nonumber \\
a_2 ~+~ a_3 ~+~ \alpha_1 &=& 2\mu ~+~ 1 \nonumber \\
a_3 ~+~ a_1 ~+~ \alpha_2 &=& 2\mu ~+~ 1 
\label{polytrg}
\end{eqnarray}
where $\alpha_i$ are the exponents of the original bare vertex of the 
underlying Lagrangian and $a_i$ are the arbitrary internal exponents. By way of
orientation, for example, one could have $\alpha_1$~$=$~$\alpha_2$~$=$~$\alpha$
and $\alpha_3$~$=$~$\beta_\sigma$ for the $\sigma \bar{\psi} \psi$ vertex of 
(\ref{lagnjluniv}). The overall function $f(\alpha_i,a_i)$ represents the 
normalization. While it is straightforward to solve the simultaneous equations 
(\ref{polytrg}) to find the $a_i$ in terms of $\alpha_i$ in general we will 
leave this to its explicit use of the construction for (\ref{lagnjluniv}).

{\begin{figure}[ht]
\begin{center}
\includegraphics[width=16cm,height=7cm]{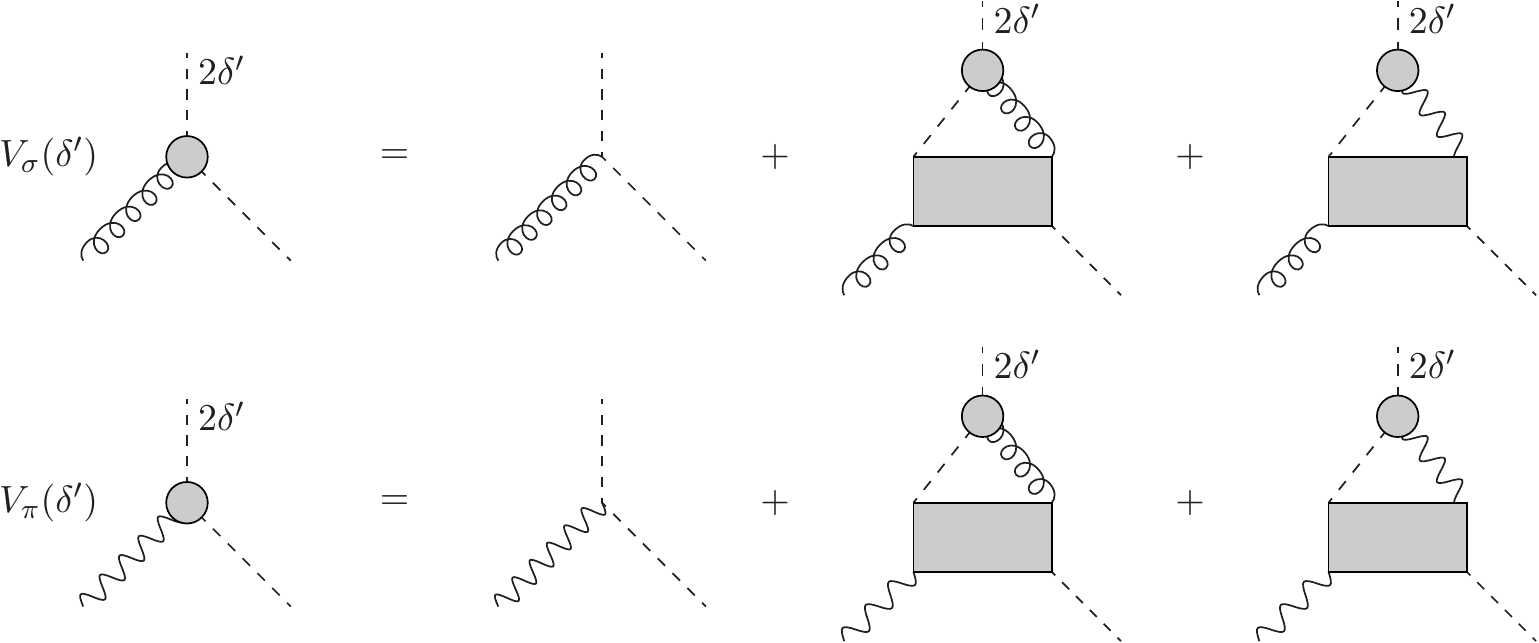}
\end{center}
\caption{$\delta^\prime$ regularized vertex functions.}
\label{figvpreg}
\end{figure}}

{\begin{figure}[ht]
\begin{center}
\includegraphics[width=13cm,height=7cm]{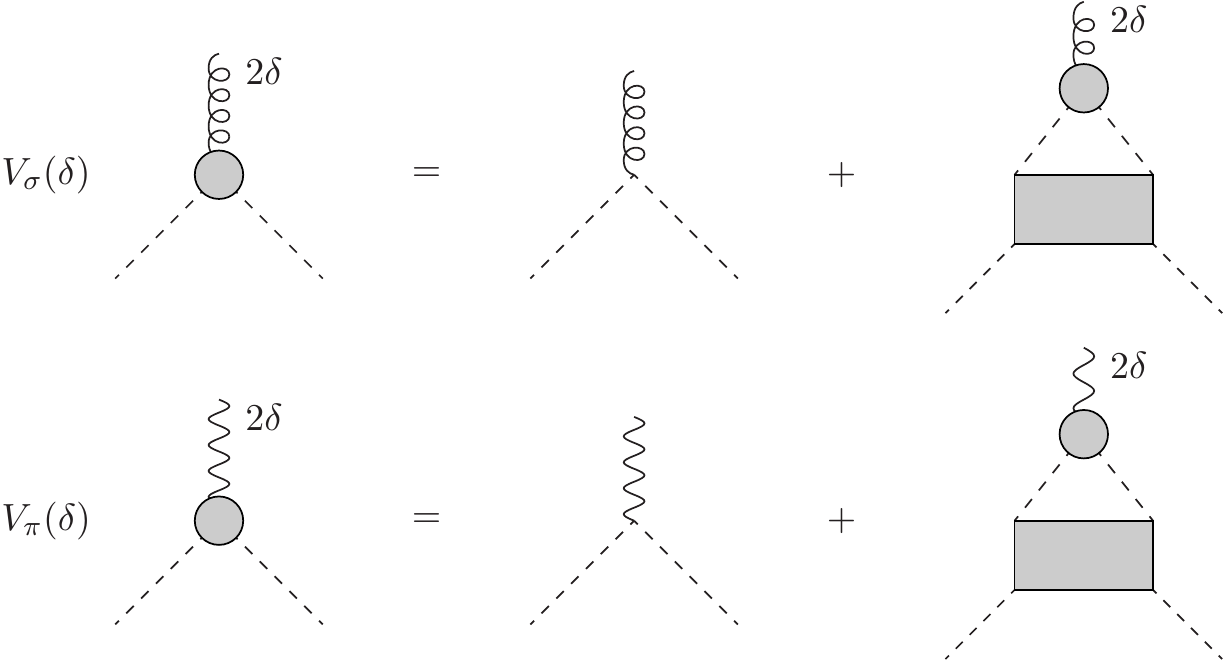}
\end{center}
\caption{$\delta$ regularized vertex functions.}
\label{figvreg}
\end{figure}}

One of the reasons for determining the various exponents for (\ref{lagnjluniv})
at $O(1/N^2)$ using the skeleton Schwinger-Dyson equations was in part to
establish $\eta$, $\chi_\sigma$ and $\chi_\pi$ at this order for the 
computation of $\eta$ at $O(1/N^3)$. Since this will use a different formalism 
knowledge of $\eta_2$ will serve as an independent check while 
$\chi_{\sigma\,2}$ and $\chi_{\pi\,2}$ are necessary to extract $\eta_3$. The 
main difference in the two formalisms rests in how the class of Feynman 
diagrams are analysed in the critical region. In the conformal bootstrap the 
key graphs are the primitive ones in the sense that not only are there no 
self-energy corrections but there are no vertex corrections. In the former case
this was because the propagators are dressed but now in the bootstrap approach 
the vertices are also dressed. So the focus is not on $2$-point functions per 
se but instead on the vertex or $3$-point functions which we denote by 
$V_\sigma$ and $V_\pi$ for the respective vertices of (\ref{lagnjluniv}). The 
method to construct the consistency equations for each here has been discussed 
previously, \cite{25,26,37}, based on the early work of \cite{29,30,44,45,46}. 
As that work involved Lagrangians with a single coupling constant we have 
repeated those derivations for (\ref{lagnjluniv}) which has two independent 
vertices and will now summarize. 

{\begin{figure}[ht]
\begin{center}
\includegraphics[width=14cm,height=8.5cm]{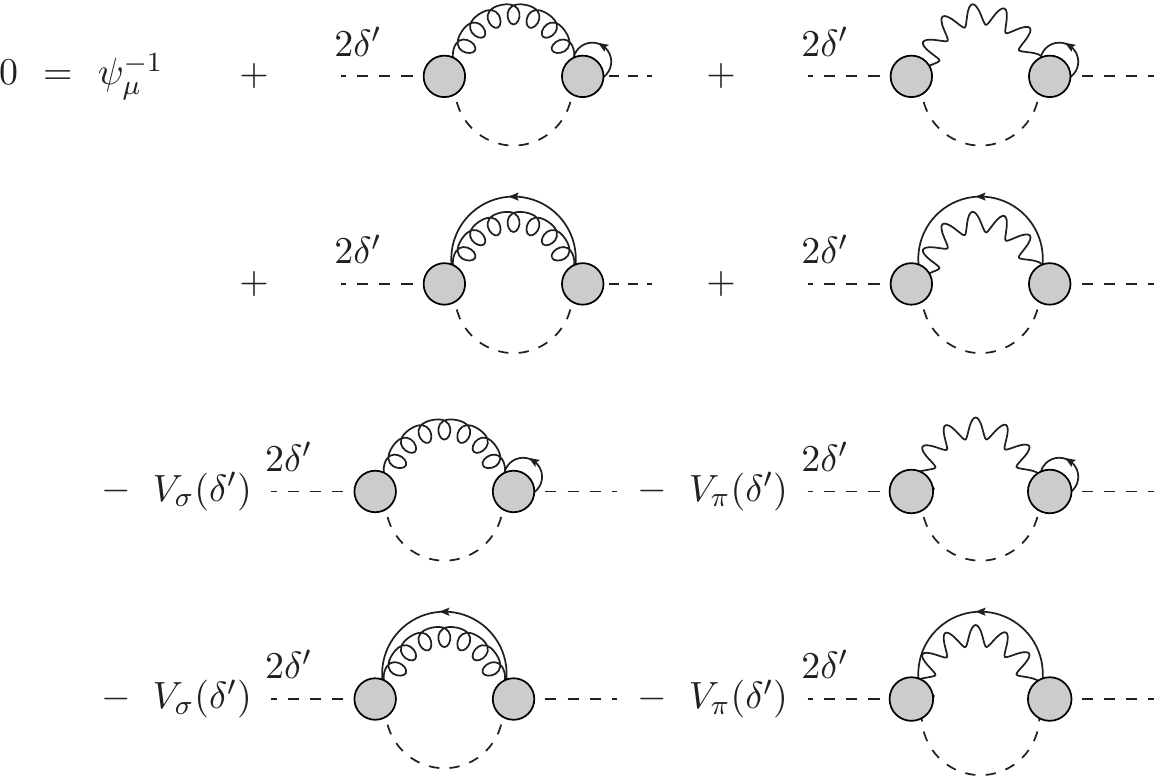}
\end{center}
\caption{$\delta^\prime$ regularized Schwinger-Dyson equation for $\psi$ 
consistency equation.}
\label{figpsidv}
\end{figure}}

The starting point is the Schwinger-Dyson equations for the three $2$-point
functions of Figures \ref{figpsicb} and \ref{figsigcb} where the blob 
represents the full vertex which will correspond to the Polyakov conformal
triangle. Clearly the graphs in each figure contain the explicit ones of 
Figures \ref{figpsi}, \ref{figsig} and \ref{figpi}. The next step in the 
derivation is to use the Schwinger-Dyson equation defining each vertex function
to replace the bare vertex in each $2$-point function of Figures \ref{figpsicb}
and \ref{figsigcb}, \cite{45,46}. These are shown in Figures \ref{figvpreg} and
\ref{figvreg} where the shaded box indicates all possible contributions to the 
respective $4$-point functions. The two relations illustrated in Figure 
\ref{figvpreg}, for example, are used to substitute for the bare vertices in 
the $\psi$ Schwinger-Dyson equation of Figure \ref{figpsicb}. In both equations
for $V_\sigma$ and $V_\pi$ one has to take account of the two interactions in 
(\ref{lagxyuniv}) rather than one as in previous Gross-Neveu bootstrap 
computations. The appearance of the regularizing parameter $\delta^\prime$ is 
due to the fact that in the formalism of \cite{45,46} there is at least one 
divergent graph in the final result for the conformal bootstrap consistency 
equation for each field. The regularization is introduced in the exponent of 
the external line where the designation of $2\delta$ and $2\delta^\prime$
appears. Unlike \cite{20} we will denote the regularizing parameters by 
$\delta$ and $\delta^\prime$ instead of $\epsilon$ and $\epsilon^\prime$ in 
order to avoid confusion with the parameter that is usually associated with the 
regulator of dimensional regularization. By contrast we recall that here the 
regularization is analytic.

{\begin{figure}[ht]
\begin{center}
\includegraphics[width=8cm,height=6cm]{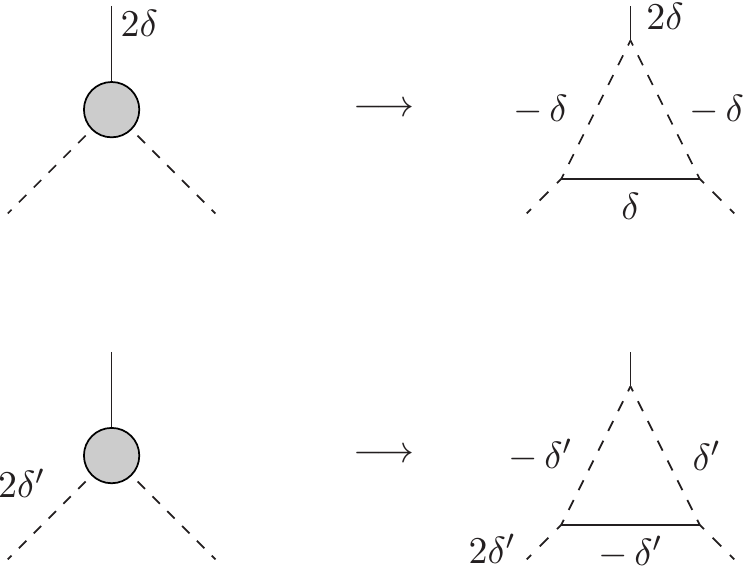}
\end{center}
\caption{Distribution of regularizing parameters $\delta$ and $\delta^\prime$
in a conformal triangle.}
\label{figctreg}
\end{figure}}

The situation after following the procedure of \cite{20,45,46} is illustrated 
in Figure \ref{figpsidv}. In that equation the Lorentz index indicates that the 
original full Schwinger-Dyson equation of Figure \ref{figpsicb} has been 
multiplied by the vector $x_\mu$, \cite{45,46}, where $x$ is the location of
one of the external vertices with the other at the origin. In following the 
manipulations to arrive at this equation in Figure \ref{figpsidv} the 
contribution from the graphs involving the $4$-point boxes have been 
eliminated. In terms of notation the directed solid line represents a 
coordinate space vector joining the endpoints of the respective full $3$-point 
vertices. At this point we divide the graphs into those with the directed line 
joining the external points of one full vertex subgraph and those where there 
is a directed line between two different full vertices. The former set are 
divergent while the latter are finite. To see this explicitly one replaces the 
full vertex by the equivalent Polyakov triangle defined by (\ref{polytrg}). As 
each of the vertices of the resulting three loop graphs are unique it is a 
simple exercise to apply the Yukawa uniqueness relation to find explicit 
expressions for all the graphs. For the graphs where the evaluation produces a 
finite expression then there is a cancellation due to the relative minus sign. 
This is because the regularized vertex functions $V_\sigma(\delta^\prime)$ and 
$V_\pi(\delta^\prime)$ are unity in the limit $\delta^\prime$~$\to$~$0$, 
\cite{20,45,46}. By contrast for the divergent graphs one has to retain the 
next term of the Taylor series in $\delta^\prime$ before taking the 
$\delta^\prime$~$\to$~$0$ limit. 

{\begin{figure}[ht]
\begin{center}
\includegraphics[width=10.5cm,height=2.5cm]{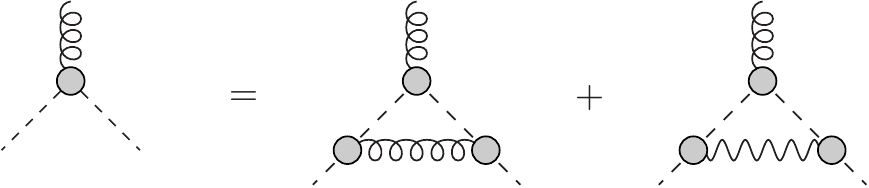}
\end{center}
\caption{Respective $O(1/N)$ graphs $V_{1\,\sigma\sigma\sigma}$ and
$V_{1\,\sigma\pi\pi}$ contributing to $V_\sigma$.}
\label{figvs1}
\end{figure}}

To assist with the evaluation we note that the distribution of the regularizing
parameters within a conformal triangle is shown in Figure \ref{figctreg} where 
these are added or subtracted to the extant internal exponents. In defining the
regularization with a shift of $2\delta$ rather than $\delta$, \cite{20}, we 
are merely avoiding the appearance of $\half\delta$ internally in the conformal 
triangle. There is no ambiguity as ultimately the regularization will be 
lifted. We note that in Figure \ref{figvpreg} and other figures we only include
the relevant argument of the vertex functions for brevity. The full dependence 
on the various exponents and parameters in fact is
\begin{equation}
V_\sigma ~=~ V_\sigma(\bar{z},\bar{y};\alpha_\psi,\beta_\sigma,\beta_\pi;\delta,
\delta^\prime) ~~,~~  
V_\pi ~=~ V_\pi(\bar{z},\bar{y};\alpha_\psi,\beta_\sigma,\beta_\pi;\delta,
\delta^\prime) ~.
\end{equation}
The parameters $\bar{z}$ and $\bar{y}$ are similar in origin to those of the
skeleton Schwinger-Dyson equations of the earlier approach in that they depend 
on the amplitude combinations $A_\psi^2 B_\sigma$ and $A_\psi^2 B_\pi$. They 
differ in value, however, due to the normalization factor in the definition of 
the conformal triangle shown in Figure \ref{figcntr}. The result of the 
exercise is the conformal bootstrap equation for $\psi$ which is 
\begin{eqnarray}
r(\alpha_\psi-1) &=& \bar{z} t_\sigma \left. 
\frac{\partial~}{\partial \delta^\prime}
V_\sigma(\bar{z},\bar{y};\alpha_\psi,\beta_\sigma,\beta_\pi;
\delta,\delta^\prime) \right|_{\delta=0,\delta^\prime=0} \nonumber \\
&& -~ C_F \bar{y} t_\pi \left. 
\frac{\partial~}{\partial \delta^\prime}
V_\pi(\bar{z},\bar{y};\alpha_\psi,\beta_\sigma,\beta_\pi;\delta,\delta^\prime)
\right|_{\delta=0,\delta^\prime=0} 
\label{cbeqnpsi}
\end{eqnarray}
where we have included the group factors associated with (\ref{lagnjluniv}) and
\begin{eqnarray}
t_\sigma ~=~ \frac{a^2(a_{\psi | \sigma}-1) a^2(\alpha_\psi-1) 
a(b_\sigma) a(\beta_\sigma)}
{(\alpha_\psi-1)^2 (a_{\psi | \sigma}-1)^2 a(\beta_\sigma-b_\sigma)} 
\nonumber \\
t_\pi ~=~ \frac{a^2(a_{\psi | \pi}-1) a^2(\alpha_\psi-1) 
a(b_\pi) a(\beta_\pi)}
{(\alpha_\psi-1)^2 (a_{\psi | \pi}-1)^2 a(\beta_\pi-b_\pi)} ~.
\end{eqnarray}
These are related to the residue with respect to $\delta^\prime$ of the
earlier divergent graphs of Figure \ref{figpsicb}. In the expression for 
$t_\sigma$ the notation is 
\begin{equation}
a_{\psi | \sigma} ~=~ \mu ~+~ \half - \half \beta_\sigma ~~~,~~~
b_\sigma ~=~ \mu ~+~ \half ~-~ \alpha_\psi ~+~ \half \beta_\sigma 
\label{cfbsig}
\end{equation}
while 
\begin{equation}
a_{\psi | \pi} ~=~ \mu ~+~ \half - \half \beta_\pi ~~~,~~~
b_\pi ~=~ \mu ~+~ \half ~-~ \alpha_\psi ~+~ \half \beta_\pi
\label{cfbpi}
\end{equation}
for $t_\pi$ which are the solutions to (\ref{polytrg}) for the internal 
exponents. In arriving at (\ref{cbeqnpsi}) we have also made use of another 
result of \cite{20,45,46} in the derivation of the bootstrap equations for the 
vertex functions which is
\begin{equation}
1 ~=~ V_\sigma(\bar{z},\bar{y};\alpha_\psi,\beta_\sigma,\beta_\pi;0,0) ~~,~~
1 ~=~ V_\pi(\bar{z},\bar{y};\alpha_\psi,\beta_\sigma,\beta_\pi;0,0) ~.
\label{cfeqn1}
\end{equation}
These reflect the sum of all the graphs with dressed propagators and vertices
contributing to the two $3$-point vertices. The $1/N$ leading order graphs for 
both $V_\sigma$ and $V_\pi$ are shown in Figures \ref{figvs1} and \ref{figvp1}
which contain the graphs of Figures \ref{figchi1} and \ref{figchi2a}. The
regularization is not necessary for these bootstrap equations since none of the
contributing graphs are divergent.

{\begin{figure}[hb]
\begin{center}
\includegraphics[width=10.5cm,height=2.5cm]{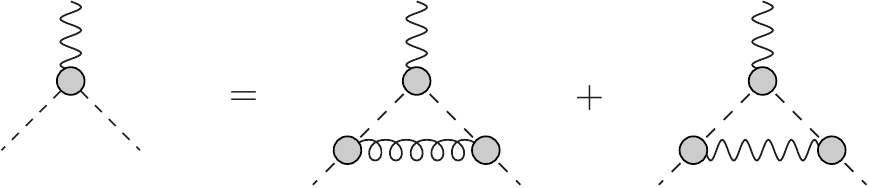}
\end{center}
\caption{Respective $O(1/N)$ graphs $V_{1\,\pi\sigma\sigma}$ and
$V_{1\,\pi\pi\pi}$ contributing to $V_\pi$.}
\label{figvp1}
\end{figure}}

The final part of the exercise in constructing the bootstrap equations is to
repeat the procedure that gave (\ref{cbeqnpsi}) for the remaining two equations
of Figure \ref{figsigcb}. The process is similar except that now in these 
graphs there is a $\delta$ regularization on the external $\sigma$ and $\pi^a$
fields. So the initial part of the derivation of the equations uses the 
regularized results of Figure \ref{figvreg} where the $4$-point boxes will
contain graphs with both $\sigma$ and $\pi^a$ fields. This is the reason for 
the simpler forms compared with Figure \ref{figvpreg}. The result of 
implementing the regularized vertex of Figure \ref{figvreg} is shown in 
Figure \ref{figsigdv} for $\sigma$ where the Lorentz index indicates that the
scaling function has been multiplied by $x_\mu$ too. The corresponding graphs
for $\pi^a$ are similar to Figure \ref{figsigdv} with the $\sigma$ external
legs replaced with $\pi^a$ fields in addition to the replacement of 
$\sigma_\mu^{-1}$ with $\pi_\mu^{-1}$. While the four graphs of Figure
\ref{figsigdv} differ from the corresponding ones of Figure \ref{figpsidv} it
is still the case that those with the directed line connecting two conformal
triangles are finite. Evaluating the remaining divergent graphs one arrives
at the respective conformal bootstrap consistency equation for $\sigma$ and
$\pi^a$ which are
\begin{eqnarray}
p(\beta_\sigma) &=& N \Nc \bar{z} t_\sigma \left. 
\frac{\partial~}{\partial \delta}
V_\sigma(\bar{z},\bar{y};\alpha_\psi,\beta_\sigma,\beta_\pi;
\delta,\delta^\prime)
\right|_{\delta=0,\delta^\prime=0} \nonumber \\
p(\beta_\pi) &=& -~ N T_F \bar{y} t_\pi \left. 
\frac{\partial~}{\partial \delta}
V_\pi(\bar{z},\bar{y};\alpha_\psi,\beta_\sigma,\beta_\pi;\delta,\delta^\prime)
\right|_{\delta=0,\delta^\prime=0}
\label{cbeqnsigpi}
\end{eqnarray}
where the same factors $t_\sigma$ and $t_\pi$ occur. Eliminating these and the
amplitudes $\bar{z}$ and $\bar{y}$ finally produces the consistency equation 
for $\eta$ which is
\begin{eqnarray}
r(\alpha_\psi-1) &=& \frac{p(\beta_\sigma)}{N \Nc} 
\left[ \left. \frac{\partial~}{\partial \delta^\prime}
V_\sigma(\bar{z},\bar{y};\alpha_\psi,\beta_\sigma,\beta_\pi;
\delta,\delta^\prime) \right| \right] 
\left[ \left. 
\frac{\partial~}{\partial \delta}
V_\sigma(\bar{z},\bar{y};\alpha_\psi,\beta_\sigma,\beta_\pi;
\delta,\delta^\prime)
\right| \right]^{-1} \nonumber \\
&& +~ \frac{C_F p(\beta_\pi)}{N T_F}
\left[ \left. 
\frac{\partial~}{\partial \delta^\prime}
V_\pi(\bar{z},\bar{y};\alpha_\psi,\beta_\sigma,\beta_\pi;\delta,\delta^\prime)
\right| \right] 
\left[ \left. 
\frac{\partial~}{\partial \delta}
V_\pi(\bar{z},\bar{y};\alpha_\psi,\beta_\sigma,\beta_\pi;\delta,\delta^\prime)
\right| \right]^{-1} ~~~
\label{cfbconeqn}
\end{eqnarray}
where the restriction indicates that both regularizations are lifted at the
end of the evaluation of the underlying graphs. At this stage (\ref{cfbconeqn})
is exact and its solution would determine $\eta$ to all orders. However to
achieve this would require the explicit values from all the contributing 
graphs.

{\begin{figure}[hb]
\begin{center}
\includegraphics[width=14cm,height=4.0cm]{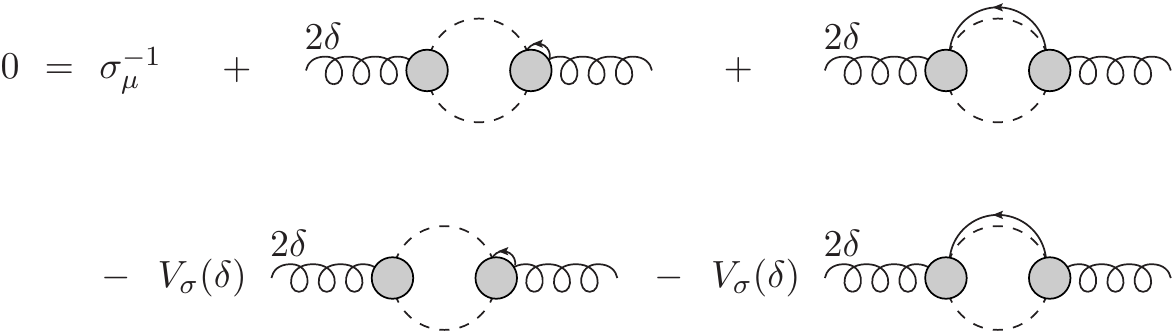}
\end{center}
\caption{$\delta$ regularized Schwinger-Dyson equation for $\sigma$ consistency
equation.}
\label{figsigdv}
\end{figure}}

The consistency equation (\ref{cfbconeqn}) can be refined though by recalling 
the observation of in \cite{20}. For the moment we consider a general situation
and denote a regularized $n$-point function by 
$\Gamma_{(n)}(\Delta_i,\delta,\delta^\prime)$ where each vertex is represented 
by a conformal triangle, the $j$th vertex has anomalous dimension 
$\chi_j$~$\equiv$~$2 \Delta_j$ and two of the external legs have $\delta$ and 
$\delta^\prime$ regularizations. Then the Green's function will have a simple 
structure. In the unregularized case it will formally be
$\Gamma(\Delta_i,0,0)/(\prod_{j=1}^n \Delta_j)$ but in the regularized case
\begin{equation}
\Gamma_{(n)}(\Delta_i,\delta,\delta^\prime) ~=~ 
\frac{\Gamma(\Delta_i,\delta,\delta^\prime)}
{(\Delta_1-\delta)(\Delta_2-\delta^\prime)(\prod_{j=3}^n \Delta_j)}
\end{equation}
where legs $1$ and $2$ are chosen to have the respective regularizations
$\delta$ and $\delta^\prime$ and $\Gamma_{(n)}(\Delta_i,\delta,\delta^\prime)$ 
is analytic in all the $\Delta_i$ and both regulators. Then
\begin{eqnarray}
\frac{\partial ~}{\partial \delta}
\Gamma_{(n)}(\Delta_i,\delta,\delta^\prime) &=& 
\frac{1}{\Delta_1} ~+~
\frac{1}{(\prod_{j=1}^n \Delta_j)} \left. \frac{\partial ~}{\partial \delta} 
\Gamma(\Delta_i,\delta,\delta^\prime) \right|_{\delta=0,\delta^\prime=0} 
\nonumber \\
\frac{\partial ~}{\partial \delta^\prime}
\Gamma_{(n)}(\Delta_i,\delta,\delta^\prime) &=& 
\frac{1}{\Delta_2} ~+~
\frac{1}{(\prod_{j=1}^n \Delta_j)} \left. 
\frac{\partial ~}{\partial \delta^\prime}
\Gamma(\Delta_i,\delta,\delta^\prime) \right|_{\delta=0,\delta^\prime=0} 
\end{eqnarray}
for instance. The first term depends only on $\Delta_1$ and $\Delta_2$
respectively since the equivalent of (\ref{cfeqn1}) has been employed and the 
relation can be written more compactly as, \cite{20},
\begin{eqnarray}
\frac{\partial ~}{\partial \delta} 
\Gamma_{(n)}(\Delta_i,\delta,\delta^\prime) &=& 
\frac{1}{\Delta_1} \left[ 1 ~+~ 
\Delta_1 \left. \frac{\partial \Gamma_{(n)}}{\partial \delta} 
\right|_{{\mbox{\footnotesize{res}}}; \, \delta=0,\delta^\prime=0} \right]
\nonumber \\
\frac{\partial ~}{\partial \delta^\prime}
\Gamma_{(n)}(\Delta_i,\delta,\delta^\prime) &=& 
\frac{1}{\Delta_2} \left[ 1 ~+~ 
\Delta_2 \left. \frac{\partial \Gamma_{(n)}}{\partial \delta^\prime} 
\right|_{{\mbox{\footnotesize{res}}}; \, \delta=0,\delta^\prime=0} \right]
\end{eqnarray}
where res indicates the contribution from the derivative of the residue of 
$\Gamma_{(n)}(\Delta_i,\delta,\delta^\prime)$ only. This was for a general 
scenario but returning to (\ref{lagnjluniv}) and applying this procedure to 
(\ref{cfbconeqn}) we have 
\begin{eqnarray}
r(\alpha_\psi-1) &=& \frac{p(\beta_\sigma)}{N \Nc} 
\left[ 1 ~+~ \Delta_\sigma 
\left. \frac{\partial V_\sigma}{\partial \delta^\prime}
\right|_{{\mbox{\footnotesize{res}}}; \, \delta=0,\delta^\prime=0} \right]
\left[ 1 ~+~ \Delta_\sigma \left. 
\frac{\partial V_\sigma}{\partial \delta}
\right|_{{\mbox{\footnotesize{res}}}; \, \delta=0,\delta^\prime=0} \right]^{-1}
\nonumber \\
&& +~ \frac{C_F p(\beta_\pi)}{N T_F}
\left[ 1 ~+~ \Delta_\pi 
\left. \frac{\partial V_\pi}{\partial \delta^\prime}
\right|_{{\mbox{\footnotesize{res}}}; \, \delta=0,\delta^\prime=0} \right]
\left[ 1 ~+~ \Delta_\pi \left. 
\frac{\partial V_\pi}{\partial \delta}
\right|_{{\mbox{\footnotesize{res}}}; \, \delta=0,\delta^\prime=0} \right]^{-1}
\label{cfeqn2}
\end{eqnarray}
where we have now set
\begin{equation}
\chi_\sigma ~=~ 2 \Delta_\sigma ~~~,~~~ \chi_\pi ~=~ 2 \Delta_\pi 
\end{equation}
and the arguments of the two vertex functions have been suppressed for brevity 
but are the same as the corresponding terms in (\ref{cfeqn2}).

{\begin{figure}[ht]
\begin{center}
\includegraphics[width=15cm,height=9.0cm]{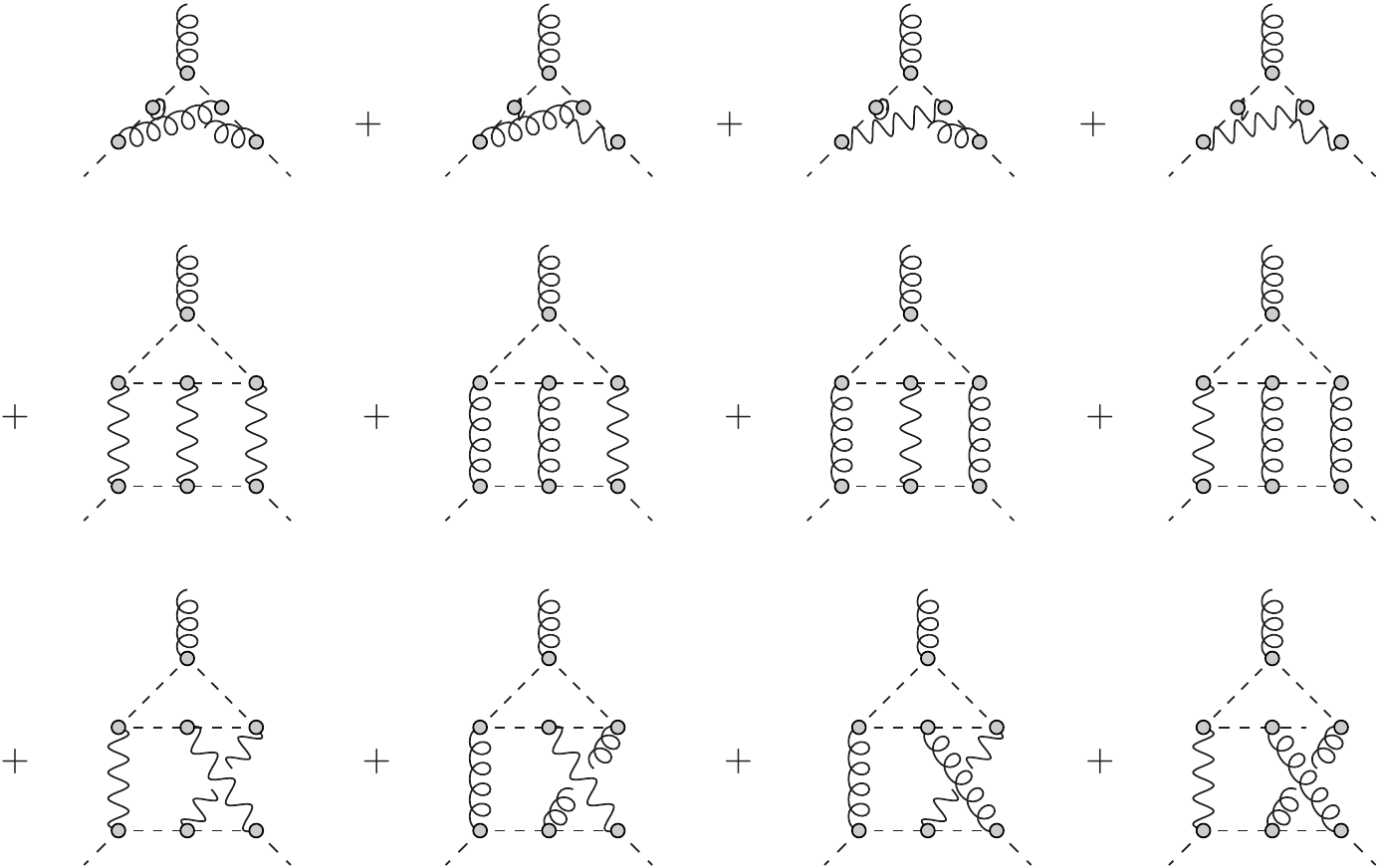}
\end{center}
\caption{$O(1/N^2)$ graphs contributing to $V_\sigma$.}
\label{figvs2}
\end{figure}}

For practical purposes it is best to expand the two vertex functions by setting
\begin{equation}
V_\sigma ~=~ \sum_{i=1}^\infty \frac{V_{\sigma \, i}}{N^i} ~~,~~
V_\pi ~=~ \sum_{i=1}^\infty \frac{V_{\pi \, i}}{N^i} 
\end{equation}
although we will only need $V_{\sigma \, 1}$, $V_{\sigma \, 2}$, $V_{\pi \, 1}$
and $V_{\pi \, 2}$ to determine $\eta_3$. If one focuses on the leading order 
large $N$ piece of (\ref{cfeqn2}) it is equivalent to that which produced 
$\eta_1$ earlier. So, for instance, a contribution to the $O(1/N^3)$ part from
the first term on  the right hand side of (\ref{cfeqn2}) can be simplified to
\begin{equation}
\Delta_{\sigma\,1} \left. \left[ 
\frac{\partial V_{\sigma\,2}}{\partial \delta^\prime} ~-~
\frac{\partial V_{\sigma\,2}}{\partial \delta} \right]
\right|_{{\mbox{\footnotesize{res}}}; \, \delta=0,\delta^\prime=0} 
\label{gamdiff}
\end{equation}
for the $V_\sigma$ vertex while the contribution from the second term involving
$V_\pi$ is similar. 

\sect{Evaluation of $\eta_3$.}

Having derived the key equation (\ref{cfeqn2}) satisfied by $\eta_3$ we now
turn to details of its determination. This entails computing the corrections to
the vertex functions of the conformal bootstrap equations (\ref{cfeqn1}) and 
(\ref{cfeqn2}). The leading order contributions are shown in Figures 
\ref{figvs1} and \ref{figvp1} and correspond to $V_{\sigma\,1}$ and
$V_{\pi\,1}$ respectively where
\begin{equation}
V_{\sigma\,1} ~=~ V_{1\,\sigma\sigma\sigma} ~+~ V_{1\,\sigma\pi\pi} ~~,~~
V_{\pi\,1} ~=~ V_{1\,\pi\sigma\sigma} ~+~ V_{1\,\pi\pi\pi} ~.
\end{equation}
At next order the graphs that comprise $V_{\sigma\,2}$ and $V_{\pi\,2}$ are 
illustrated in Figures \ref{figvs2} and \ref{figvp2} and are clearly primitive.
While the leading order $1/N$ terms of both $V_{\sigma\,1}$ and $V_{\pi\,1}$ 
are needed to verify $\eta_2$ the next term in the series is required for 
$\eta_3$. This $N$ dependence arises through the exponents $\alpha_\psi$, 
$\beta_\sigma$ and $\beta_\pi$. Therefore what are sometimes termed $3$-gamma 
graphs, which are illustrated in Figures \ref{figvs1} and \ref{figvp1}, have to
be computed. This is achieved via the method developed in \cite{20} using the 
properties of the Polyakov conformal triangle.

{\begin{figure}[ht]
\begin{center}
\includegraphics[width=15cm,height=9.0cm]{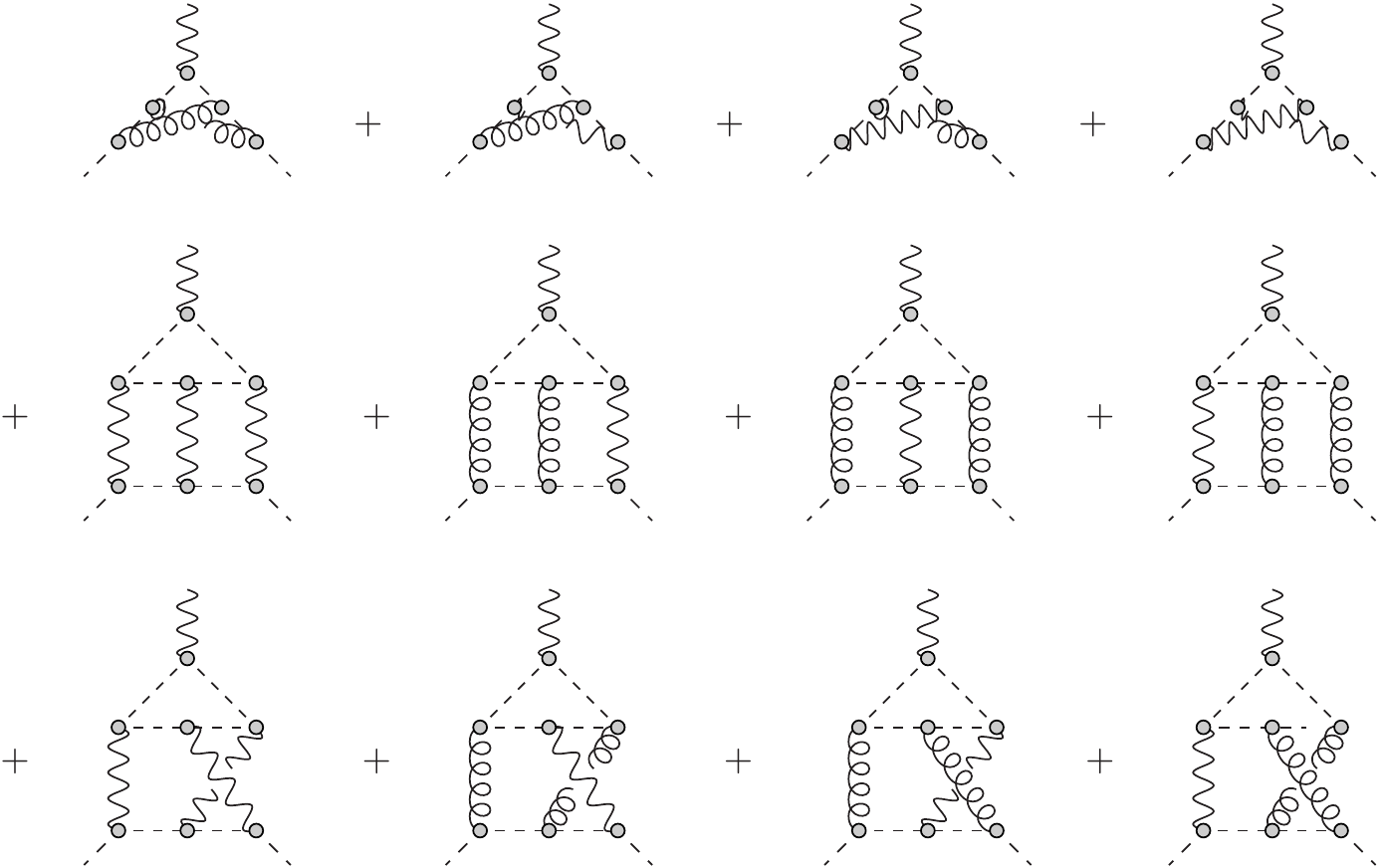}
\end{center}
\caption{$O(1/N^2)$ graphs contributing to $V_\pi$.}
\label{figvp2}
\end{figure}}

For instance, the explicit designation of the exponents on the lines of both 
$V_{1\,\sigma\sigma\sigma}$ and $V_{1\,\sigma\pi\pi}$ are shown in Figures 
\ref{gamsss} and \ref{gamspp} where
\begin{equation}
\gamma ~=~ 1 ~-~ \eta 
\end{equation}
is set for shorthand due to $\beta_\sigma$ and $\beta_\pi$ differing only in 
the piece involving the respective vertex anomalous dimensions. The factor of
$2$ associated with $\Delta_\sigma$ and $\Delta_\pi$ is a consequence of
(\ref{cfbsig}) and (\ref{cfbpi}). The graph of Figure \ref{gamsss} was 
evaluated in \cite{27}. If we set
\begin{equation}
V_{1 \, \sigma \sigma \sigma} ~=~ 
-~ \frac{Q^3}{\Delta_\sigma(\Delta_\sigma-\delta)(\Delta_\sigma-\delta^\prime)}
\exp [ F_{\sigma\sigma\sigma}(\delta, \delta^\prime,\Delta_\sigma)]
\end{equation}
which is consistent with the general form of $\Gamma_{(n)}$ then we recall that 
\begin{eqnarray}
F_{\sigma\sigma\sigma}(\delta,\delta^\prime,\Delta_\sigma) &=& 
\left[ 5 B_\gamma - 2 B_{\alpha_\psi-1} - 3 B_0 
- \frac{2}{(\alpha_\psi-1)}\right]\Delta_\sigma 
- \left[ B_\gamma - B_0 \right] \delta 
\nonumber \\
&& 
+~ \left[ B_0 - B_{\alpha_\psi-1} - \frac{1}{(\alpha_\psi-1)} \right] 
\delta^\prime
+ \left[ C_{\alpha_\psi-1} - \frac{1}{(\alpha_\psi-1)} \right] 
\delta \delta^\prime
\nonumber \\
&& 
+~ \left[ C_\gamma + C_0 - 2 C_{\alpha_\psi-1} + \frac{2}{(\alpha_\psi-1)^2} 
\right] \Delta_\sigma\delta 
\nonumber \\
&& 
+~ \frac{1}{2} \left[ \frac{1}{(\alpha_\psi-1)^2} - C_{\alpha_\psi-1} 
- C_0 \right] {\delta^\prime}^2 
\nonumber \\
&& 
+~ \left[ C_0 - C_\gamma - 2 C_{\alpha_\psi-1} + \frac{2}{(\alpha_\psi-1)^2} 
\right] \Delta_\sigma \delta^\prime
- \frac{1}{2} \left[ C_\gamma + C_0 \right] \delta^2 
\nonumber \\
&& 
+~ \left[ C_{\alpha_\psi-1} - \frac{7}{2} C_\gamma - \frac{3}{2} C_0
- \frac{1}{(\alpha_\psi-1)^2} \right]\Delta_\sigma^2 
\label{v1sss}
\end{eqnarray}
with the shorthand notation, \cite{20},
\begin{eqnarray}
B_z &=& \psi(\mu-z) ~+~ \psi(z) ~~~,~~~ B_0 ~=~ \psi(1) ~+~ \psi(\mu)
\nonumber \\
C_z &=& \psi^\prime(z) ~-~ \psi^\prime(\mu-z) ~~~,~~~ 
C_0 ~=~ \psi^\prime(\mu) ~-~ \psi^\prime(1) 
\end{eqnarray}
and
\begin{equation}
Q ~=~ - \, \frac{\pi^{2\mu} a^2(\alpha_\psi-1) a(\gamma)}
{(\alpha_\psi-1)^2\Gamma(\mu)} ~.
\end{equation}
We note that in \cite{27} not all the terms quadratic in combinations of the 
parameters $\delta$ or $\delta^\prime$ were recorded. We have included them 
here for completeness and as an aid to an interested reader although they do 
not contribute to the consistency equation. 

{\begin{figure}[ht]
\begin{center}
\includegraphics[width=11cm,height=7.5cm]{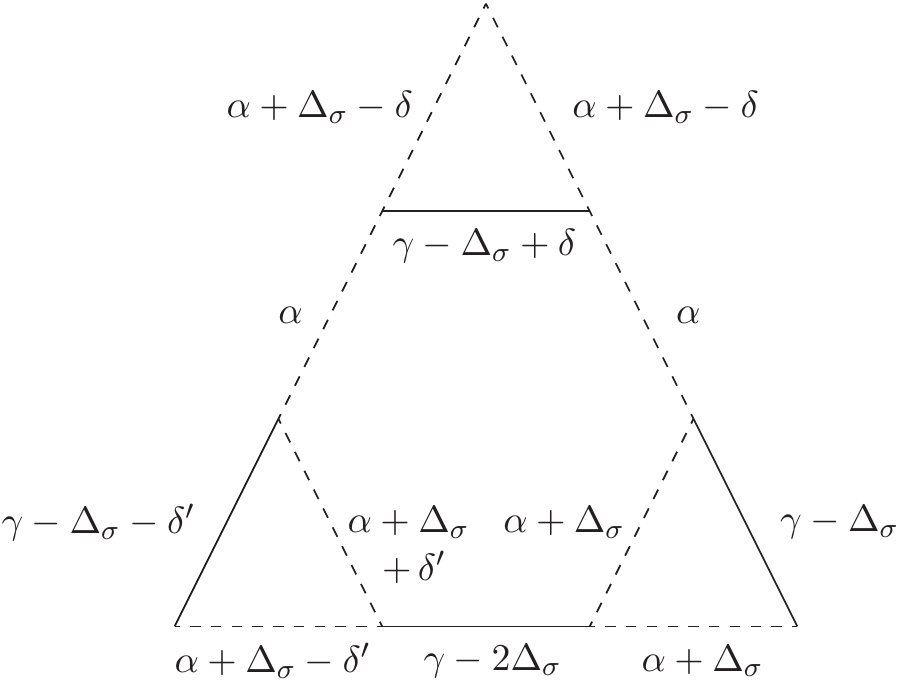}
\end{center}
\caption{Regularized one loop $3$-gamma graph denoted by
$V_{1\,\sigma\sigma\sigma}$ contributing to $V_\sigma$ containing a $\sigma$
propagator and including the conformal triangles.}
\label{gamsss}
\end{figure}}

We have rederived (\ref{v1sss}) that appeared in \cite{27} for 
$V_{1 \, \sigma \sigma \sigma}$ in order to extend it for 
$V_{1 \, \sigma \pi \pi}$. The latter is more general than the former since it 
will depend on $\chi_\pi$ as well as $\chi_\sigma$. This leads to a check in 
that formally taking the limit $\chi_\pi$~$\to$~$\chi_\sigma$ in 
$V_{1 \, \sigma \pi \pi}$ should produce the expression for 
$V_{1 \, \sigma \sigma \sigma}$. If we define
\begin{equation}
\Gamma_{1 \, \sigma \pi \pi} ~=~ 
-~ \frac{Q^3}{\Delta_\pi(\Delta_\sigma-\delta)(\Delta_\pi-\delta^\prime)}
\exp [ F_{\sigma\pi\pi}(\delta, \delta^\prime,\Delta_\sigma,\Delta_\pi)]
\end{equation}
and repeat the procedure that resulted in (\ref{v1sss}) we find
\begin{eqnarray}
F_{\sigma\pi\pi}(\delta, \delta^\prime,\Delta_\sigma,\Delta_\pi) &=&
\left[ B_0 - B_\gamma \right] \delta
+ \left[ B_0 - B_{\alpha_\psi-1} - \frac{1}{(\alpha_\psi-1)} \right] 
\delta^\prime
+ \left[ B_\gamma - B_0 \right] \Delta_\sigma
\nonumber \\
&& 
+~ \left[ 4 B_\gamma - 2 B_0 - 2 B_{\alpha_\psi-1} - \frac{2}{(\alpha_\psi-1)} 
\right] \Delta_\pi
- \frac{1}{2} \left[ C_\gamma + C_0 \right] \Delta_\sigma^2
\nonumber \\
&& 
+~ \left[ \frac{1}{(\alpha_\psi-1)^2} - C_{\alpha_\psi-1} - 3 C_\gamma - C_0 
\right] \Delta_\pi^2
+ \left[ C_\gamma + C_0 \right] \delta \Delta_\sigma
\nonumber \\
&& 
+~ 2 \left[ C_{\alpha_\psi-1} - \frac{1}{(\alpha_\psi-1)^2} \right]
\Delta_\sigma \Delta_\pi
+ \left[ \frac{1}{(\alpha_\psi-1)^2} - C_{\alpha_\psi-1} \right]
\delta^\prime \Delta_\sigma
\nonumber \\
&& 
+~ \left[ C_0 - C_{\alpha_\psi-1} - C_\gamma + \frac{1}{(\alpha_\psi-1)^2} 
\right] \delta^\prime \Delta_\pi
- \frac{1}{2} \left[ C_\gamma + C_0 \right] \delta^2
\nonumber \\
&& 
+~ \frac{1}{2} \left[ \frac{1}{(\alpha_\psi-1)^2} - C_{\alpha_\psi-1} - C_0 
\right] {\delta^\prime}^2
+ 2 \left[ \frac{1}{(\alpha_\psi-1)^2} - C_{\alpha_\psi-1} \right] 
\delta \Delta_\pi
\nonumber \\
&& 
+~ \left[ C_{\alpha_\psi-1} - \frac{1}{(\alpha_\psi-1)^2} \right]
\delta \delta^\prime ~.
\label{v1spp}
\end{eqnarray}
Clearly this is consistent with (\ref{v1sss}) since
$F_{\sigma\pi\pi}(\delta, \delta^\prime,\Delta_\sigma,
\Delta_\sigma)$~$=$~$F_{\sigma\sigma\sigma}
(\delta, \delta^\prime,\Delta_\sigma)$. These two expressions are sufficient to
calculate the contribution to (\ref{cfeqn2}) from $V_{\sigma\,1}$. For the 
second term of (\ref{cfeqn2}) involving $V_{\pi\,1}$ the expressions for the
corresponding $3$-gamma graphs can be deduced from those of $V_{\sigma\,1}$ by
formally swapping $\Delta_\sigma$ and $\Delta_\pi$ in (\ref{v1sss}) and
(\ref{v1spp}). 

{\begin{figure}[ht]
\begin{center}
\includegraphics[width=11cm,height=7.5cm]{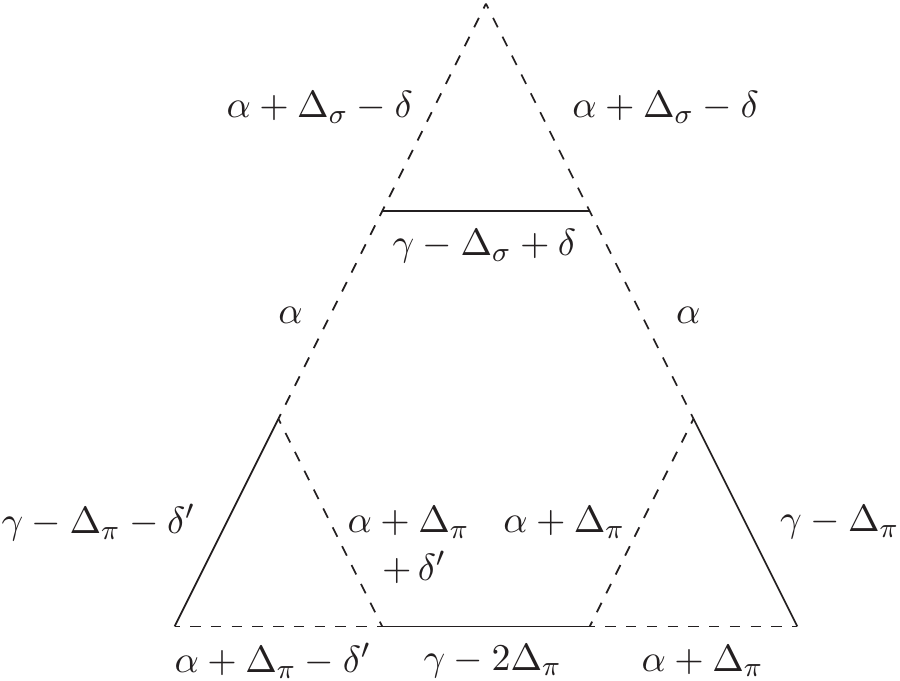}
\end{center}
\caption{Regularized one loop $3$-gamma graph denoted by
$V_{1\,\sigma\pi\pi}$ contributing to $V_\sigma$ containing a $\pi$ propagator
and including the conformal triangles.}
\label{gamspp}
\end{figure}}

To complete the evaluation of $\eta_3$ requires the contribution to
(\ref{cfeqn2}) from $V_{\sigma\,2}$ and $V_{\pi\,2}$. The values of the four
topologies underlying the graphs in Figures \ref{figvs2} and \ref{figvp2} have
been computed previously, \cite{25,26,27}. Strictly though it is the 
combination of (\ref{gamdiff}) and the $\pi^a$ counterpart that have been
calculated. The values of the individual terms cannot be extracted as the
underlying three loop integral does not have unique vertices to allow the
integration to proceed. While it would appear from the figures that there are 
three distinct topologies this would be the case if there was no 
regularization. With non-zero $\delta$ and $\delta^\prime$ through specified 
external legs there are two independent contributions from the three loop 
non-planar graphs. Indeed the contribution to (\ref{cfeqn1}) from these
non-planar graphs is the same but not for (\ref{cfeqn2}). For each of the
graphs of Figures \ref{gamsss} and \ref{gamspp} we have computed the associated
group factor by again using the {\tt color.h} package based on \cite{42}. The 
higher rank colour Casimir will arise in the graphs of Figure \ref{figvp2} 
where four $\pi^a$ fields connect to the fermion loop.

The next stage is to assemble all the components contributing to (\ref{cfeqn1})
and (\ref{cfeqn2}). The purpose of the two equations of (\ref{cfeqn1}) is to
fix the values of $\bar{z}$ and $\bar{y}$ at $O(1/N^2)$. While these were
eliminated to arrive at (\ref{cfeqn2}) combinations of them are present as
factors in each of the graphs of Figures \ref{figvs1}, \ref{figvp1},
\ref{figvs2} and \ref{figvp2}. However we can now illuminate the difficulties 
with a direct evaluation in the $U(1)$ case. The formal solution for $\bar{z}$ 
in (\ref{lagnjluniv}) at leading order is
\begin{equation}
\bar{z}_1 ~=~ \left[ [2 C_F - C_A] \Delta_{\sigma\,1}^3
~+~ 2 C_F \Delta_{\sigma\,1}^2 \Delta_{\pi\,1} \right] \frac{1}{C_A Q^3} ~.
\end{equation}
While $C_A$ vanishes in the abelian limit producing a singularity, when the
leading order values for the vertex dimensions are included these are
proportional to $C_A$. So that overall the leading order variable $\bar{z}$ is
finite in the abelian limit. For a direct evaluation one could not determine
$\bar{z}_1$ since it would equate to the mathematically ill-defined quantity
of $0/0$. The situation for $\bar{y}_1$ is similar. It was this that forced us 
to consider the more general Lagrangian. For (\ref{lagnjluniv}) we find
\begin{eqnarray}
\bar{z}_1 &=& -~ \frac{(C_F \Nc - T_F)^2 \Gamma^3(2\mu-1)}
{\Gamma^3(\mu-1)\Gamma^3(1-\mu) \Nc^3 T_F^3} \nonumber \\
\bar{y}_1 &=& -~ \frac{(2 C_F \Nc - 2 T_F - C_A \Nc)^2 \Gamma^3(2\mu-1)}
{4\Gamma^3(\mu-1)\Gamma^3(1-\mu) \Nc^3 T_F^3}
\end{eqnarray}
for instance which is finite in the abelian limit. With these values and those 
for the next order we ultimately arrive at our main result for 
(\ref{lagxyuniv}) which is
\begin{eqnarray}
\eta_3 &=& \left[
\left[ \mu^2 (\mu+8) \frac{d_F^{abcd} d_F^{abcd} \Nc^2}{T_F}
+ \frac{1}{2} \mu (3 \mu+1) C_A C_F T_F \Nc^2
- \frac{1}{2} \mu (5 \mu-1) C_F^2 C_A \Nc^3
\right. \right. \nonumber \\
&& \left. \left. ~~~
- \frac{1}{24} \mu^2 (\mu-4) C_F C_A^2 \Nc^3
+ (\mu^3+10 \mu^2+\mu-1) T_F^3
- (2 \mu^3+22 \mu^2-7 \mu+3) C_F \Nc T_F^2
\right. \right. \nonumber \\
&& \left. \left. ~~~
- (6 \mu^2-7 \mu+3) C_F^2 \Nc^2 T_F
+ (2 \mu-1) (\mu+1) C_F^3 \Nc^3 \right]
\frac{\Theta(\mu)}{4[\mu-1]^2}
\right. \nonumber \\
&& \left. ~
- \left[
4 \mu (2 \mu-1) C_F^2 C_A \Nc^3
- 4 \mu C_A C_F T_F \Nc^2
- \mu^2 C_F C_A^2 \Nc^3
- 4 (2 \mu-1)^2 T_F^3
\right. \right. \nonumber \\
&& \left. \left. ~~~~~
+ 4 (4 \mu-3) C_F \Nc T_F^2
+ 4 (4 \mu-3) C_F^2 \Nc^2 T_F
- 4 (2 \mu-1)^2 C_F^3 \Nc^3 \right]
\frac{\Phi(\mu)}{8[\mu-1]^2}
\right. \nonumber \\
&& \left. ~
+ \left[ \frac{3}{2} T_F^3 - 3 C_F \Nc T_F^2 - \frac{1}{16} C_A^2 C_F \Nc^3
+ \frac{3}{2} \frac{d_F^{abcd} d_F^{abcd} \Nc^2}{T_F} \right]
\left[ \Theta(\mu) + \frac{1}{[\mu-1]^2} \right] \frac{\mu^2\Xi(\mu)}{[\mu-1]}
\right. \nonumber \\
&& \left. ~
+ \left[ 24 d_F^{abcd} d_F^{abcd} \Nc^2 + 24 T_F^4 - 48 C_F \Nc T_F^3 
- C_A^2 C_F \Nc^3 T_F \right]
\frac{\mu^2 \Theta(\mu) \Psi(\mu)}{8[\mu-1]T_F}
\right. \nonumber \\
&& \left. ~
- \left[
4 (4 \mu-3) C_F \Nc T_F^2
- 4 \mu C_A C_F T_F \Nc^2
- \mu^2 C_F C_A^2 \Nc^3
- 4 (2 \mu-1)^2 T_F^3
\right. \right. \nonumber \\
&& \left. \left. ~~~~~
+ 4 \mu (2 \mu-1) C_F^2 C_A \Nc^3
+ 4 (4 \mu-3) C_F^2 \Nc^2 T_F
- 4 (2 \mu-1)^2 C_F^3 \Nc^3 \right]
\frac{3 \Psi^2(\mu)}{8[\mu-1]^2}
\right. \nonumber \\
&& \left. ~
+ \left[
\frac{(24 \mu^2-52 \mu+9)}{2\mu[\mu-1]^2} C_F^2 \Nc^2 T_F
- \frac{(12 \mu^2-23 \mu+2)}{4[\mu-1]^3} C_A C_F T_F \Nc^2
\right. \right. \nonumber \\
&& \left. \left. ~~~~~
- \frac{(19 \mu-2) (2 \mu-1)}{4[\mu-1]^3} C_F^2 C_A \Nc^3
- \frac{(2 \mu^5-3 \mu^4-46 \mu^3+56 \mu^2-23 \mu+3)}{2\mu[\mu-1]^3} T_F^3
\right. \right. \nonumber \\
&& \left. \left. ~~~~~
- \frac{(12 \mu^5-34 \mu^4+20 \mu^3+76 \mu^2-61 \mu+9)}{2\mu[\mu-1]^3} 
C_F \Nc T_F^2
- \frac{\mu^2 (\mu+2) (\mu-5)}{12[\mu-1]^3} C_F C_A^2 \Nc^3
\right. \right. \nonumber \\
&& \left. \left. ~~~~~
+ \frac{(11 \mu-3) (2 \mu-1)^2}{2\mu[\mu-1]^3} C_F^3 \Nc^3
- \frac{\mu^2 (2 \mu+1) (\mu-2)}{2[\mu-1]^3} \frac{d_F^{abcd} d_F^{abcd} \Nc^2}
{T_F} \right] \Psi(\mu)
\right. \nonumber \\
&& \left. ~
- \frac{3(11 \mu^2-12 \mu+2)}{8[\mu-1]^4} C_A C_F T_F \Nc^2
- \frac{\mu^2 (8 \mu^3-10 \mu^2-17 \mu-32)}{96[\mu-1]^4} C_F C_A^2 \Nc^3
\right. \nonumber \\
&& \left. ~
- \frac{3(4 \mu-1) (2 \mu-1)}{4[\mu-1]^4} C_F^2 C_A \Nc^3
- \frac{(2 \mu-1) (\mu^6-\mu^4-18 \mu^3+21 \mu^2-8 \mu+1)}{2\mu^2[\mu-1]^4} 
T_F^3
\right. \nonumber \\
&& \left. ~
- \frac{(12 \mu^7-14 \mu^6-24 \mu^5+11 \mu^4+108 \mu^3-99 \mu^2+30 \mu-3)}
{2\mu^2[\mu-1]^4} C_F \Nc T_F^2
\right. \nonumber \\
&& \left. ~
+ \frac{3(15 \mu^4-36 \mu^3+33 \mu^2-10 \mu+1)}{2\mu^2[\mu-1]^4} 
C_F^2 \Nc^2 T_F
+ \frac{(3 \mu-1)^2 (2 \mu-1)^2}{2\mu^2[\mu-1]^4} C_F^3 \Nc^3
\right. \nonumber \\
&& \left. ~
- \frac{\mu^2 (\mu+1) (2 \mu-1)}{2[\mu-1]^3} 
\frac{d_F^{abcd} d_F^{abcd} \Nc^2}{T_F}
\right] \frac{\eta_1^3}{[T_F+C_F \Nc]^3}
\label{njleta3}
\end{eqnarray}
which is considerably more involved than the other exponents. We note that
electronic versions of (\ref{njleta3}) and all the large $N$ exponents of
(\ref{lagnjluniv}) are available in an attached data file.

Like the $\eta_3$ result of \cite{20} (\ref{njleta3}) contains the function 
$\Xi(\mu)$ which is related to the function $I(\mu)$ of that paper via 
\begin{equation}
I(\mu) ~=~ -~ \frac{2}{3(\mu-1)} ~+~ \Xi(\mu) ~.
\end{equation}
The $\epsilon$ expansion of $I(\mu)$ around $d$~$=$~$D$~$-$~$2\epsilon$ is
known for integer $D$. In the case of $D$~$=$~$2$ and $4$ this was derived in
\cite{47} and built on the earlier expansion provided in \cite{20}. When 
$D$~$=$~$3$ the expansion was provided in \cite{48} where the leading order 
term is 
\begin{equation}
I(\threehalves) ~=~ 2 \ln 2 ~-~ \frac{21}{\pi^2} \zeta(3)
\end{equation}
which was derived in \cite{20} using uniqueness methods. So, for instance, we
can record the first three terms of the large $N$ expansion of each exponent in
three dimensions for (\ref{lagnjluniv}) which are
\begin{eqnarray}
\left. \eta \right|_{d=3} &=& \frac{8[C_F \Nc+ T_F]}{3\pi^2 \Nc T_F N} ~+~
\left[ 28 C_F^2 \Nc^2 - 9 C_A C_F \Nc^2 - 16 C_F \Nc T_F + 28 T_F^2 \right]
\frac{64}{27 \pi^4 \Nc^2 T_F^2 N} \nonumber \\
&& +~
\left[ 
2592 \pi^2 \ln(2) d_F^{abcd} d_F^{abcd} \Nc^2 
- 108 \pi^2 \ln(2) C_A^2 C_F \Nc^3 T_F - 5184 \pi^2 \ln(2) C_F \Nc T_F^3 
\right. \nonumber \\
&& \left. ~~~~
+ 2592 \pi^2 \ln(2) T_F^4 + 1134 \zeta(3) C_A^2 C_F \Nc^3 T_F 
+ 54432 \zeta(3) C_F \Nc T_F^3 
\right. \nonumber \\
&& \left. ~~~~
- 27216 \zeta(3) d_F^{abcd} d_F^{abcd} \Nc^2 - 27216 \zeta(3) T_F^4 
- 351 \pi^2 C_A^2 C_F \Nc^3 T_F 
\right. \nonumber \\
&& \left. ~~~~
+ 2916 C_A^2 C_F \Nc^3 T_F 
+ 1368 \pi^2 C_A C_F^2 \Nc^3 T_F - 17568 C_A C_F^2 \Nc^3 T_F 
\right. \nonumber \\
&& \left. ~~~~
- 360 \pi^2 C_A C_F \Nc^2 T_F^2 
- 35712 C_A C_F \Nc^2 T_F^2 
- 2112 \pi^2 C_F^3 \Nc^3 T_F + 33536 C_F^3 \Nc^3 T_F 
\right. \nonumber \\
&& \left. ~~~~
+ 1152 \pi^2 C_F^2 \Nc^2 T_F^2 
+ 94848 C_F^2 \Nc^2 T_F^2 
- 5328 \pi^2 C_F \Nc T_F^3 - 65856 C_F \Nc T_F^3 
\right. \nonumber \\
&& \left. ~~~~
+ 3240 \pi^2 d_F^{abcd} d_F^{abcd} \Nc^2 
- 54432 d_F^{abcd} d_F^{abcd} \Nc^2 
+ 1128 \pi^2 T_F^4 
\right. \nonumber \\
&& \left. ~~~~
- 20896 T_F^4 \right] 
\frac{8}{243 \pi^6 \Nc^3 T_F^4 N^3} ~+~ O \left( \frac{1}{N^4} \right) 
\nonumber \\
\left. \beta_\sigma \right|_{d=3} &=&
1 ~+~ \frac{32 [C_F \Nc-2 T_F]}{3\pi^2 \Nc T_F N} \nonumber \\
&& +~ \frac{64 [ 45 C_A C_F \Nc^2 - 92 C_F^2 \Nc^2 + 27 \pi^2 C_F \Nc T_F 
- 76 C_F \Nc T_F - 27 \pi^2 T_F^2 + 304 T_F^2 ]}{27 \pi^4 \Nc^2 T_F^2 N^2}
\nonumber \\
&& +~ O \left( \frac{1}{N^3} \right) \nonumber \\
\left. \beta_\pi \right|_{d=3} &=&
1 ~+~ \frac{8 [3 C_A \Nc - 8 C_F \Nc + 4 T_F]}{3\pi^2 \Nc T_F N} \nonumber \\
&& +~
\left[ 9 \pi^2 C_A^2 C_F \Nc^2 T_F - 252 C_A^2 C_F \Nc^2 T_F 
+ 864 C_A C_F^2 \Nc^2 T_F + 936 C_A C_F \Nc T_F^2
\right. \nonumber \\
&& \left. ~~~~
- 1024 C_F^3 \Nc^2 T_F - 2336 C_F^2 \Nc T_F^2 + 216 \pi^2 C_F T_F^3 
+ 992 C_F T_F^3 
\right. \nonumber \\
&& \left. ~~~~
- 216 \pi^2 d_F^{abcd} d_F^{abcd} \Nc 
+ 3456 d_F^{abcd} d_F^{abcd} \Nc \right]
\frac{8}{27 \pi^4 C_F \Nc^2 T_F^3 N^2} ~+~ O \left( \frac{1}{N^3} \right)
\nonumber \\
\left. \frac{1}{\nu} \right|_{d=3} &=&
1 ~-~ \frac{32 [C_F \Nc+ T_F]}{3\pi^2 \Nc T_F N} \nonumber \\
&& +~ 
\left[ 324 C_A^2 C_F \Nc^3 T_F - 27 \pi^2 C_A^2 C_F \Nc^3 T_F 
+ 108 \pi^2 C_A C_F^2 \Nc^3 T_F 
\right. \nonumber \\
&& \left. ~~~~
- 1260 C_A C_F^2 \Nc^3 T_F 
- 72 C_A C_F \Nc^2 T_F^2 - 108 \pi^2 C_F^3 \Nc^3 T_F
+ 1264 C_F^3 \Nc^3 T_F 
\right. \nonumber \\
&& \left. ~~~~
+ 624 C_F^2 \Nc^2 T_F^2 + 108 \pi^2 C_F \Nc T_F^3 
- 240 C_F \Nc T_F^3 + 162 \pi^2 d_F^{abcd} d_F^{abcd} \Nc^2 
\right. \nonumber \\
&& \left. ~~~~
+ 54 \pi^2 T_F^4
+ 1264 T_F^4 \right] \frac{32}{27 \pi^4 \Nc^2 T_F^3 [C_F \Nc + T_F] N^2} ~+~ 
O \left( \frac{1}{N^3} \right) 
\end{eqnarray}
from which we will be able to deduce the analogous expressions for the chiral
XY model. Finally we note that (\ref{njleta3}) agrees with known expressions in
the limit of (\ref{gnlim}). 

{\begin{table}[ht]
\begin{center}
\begin{tabular}{|c|c||c|c|c|c|}
\hline
Exponent & Pad\'{e} & $N_t ~=~ 2$ & $N_t ~=~ 4$ & $N_t ~=~ 6$ & $N_t ~=~ 8$ \\
\hline
$\eta^{\mbox{\footnotesize{XY}}}$ 
       & [1,1] & 0.087176 & 0.038058 & 0.024343 & 0.017894 \\ 
       & [1,2] & 0.078821 & 0.037198 & 0.024104 & 0.017797 \\ 
\hline
$\eta_\phi^{\mbox{\footnotesize{XY}}}$
            & [0,1] & 0.880984 & 0.936726 & 0.956909 & 0.967330 \\ 
            & [1,1] & 0.902301 & 0.943303 & 0.960064 & 0.969176 \\ 
            & [0,2] & 0.907741 & 0.944124 & 0.960325 & 0.969290 \\ 
\hline
$\frac{1}{\nu^{\mbox{\footnotesize{XY}}}}$ & [0,1]
            & 0.787284 & 0.880984 & 0.917378 & 0.936726 \\ 
            & [1,1] & 0.902616 & 0.928416 & 0.943409 & 0.953210 \\ 
\hline
\end{tabular}
\end{center}
\begin{center}
{Table $1$. Pad\'{e} approximants for exponents in chiral XY model for 
$N_t$~$=$~$2$, $4$, $6$ and $8$.}
\end{center}
\end{table}}

{\begin{table}[ht]
\begin{center}
\begin{tabular}{|c|c||c|c|}
\hline
Exponent & Reference & Method & $N_t ~=~ 2$ \\
\hline
$\eta^{\mbox{\footnotesize{XY}}}$ 
       & \cite{15} & $[2,2]$ Pad\'{e} & 0.117 \\ 
       & \cite{15} & $[3,1]$ Pad\'{e} & 0.108 \\ 
       & \cite{11} & functional RG & 0.062 \\ 
\hline
$\eta_\phi^{\mbox{\footnotesize{XY}}}$
       & \cite{15} & $[2,2]$ Pad\'{e} & 0.810 \\ 
       & \cite{15} & $[3,1]$ Pad\'{e} & 0.788 \\ 
       & \cite{11} & functional RG & 0.88 \\ 
\hline
$\frac{1}{\nu^{\mbox{\footnotesize{XY}}}}$ 
       & \cite{15} & $[2,2]$ Pad\'{e} & 0.840 \\ 
       & \cite{15} & $[3,1]$ Pad\'{e} & 0.841 \\ 
       & \cite{11} & functional RG & 0.862 \\ 
\hline
\end{tabular}
\end{center}
\begin{center}
{Table $2$. Exponent estimates in chiral XY model for $N_t$~$=$~$2$ from 
$[2,2]$ and $[3,1]$ Pad\'{e} approximants of four dimensional $\epsilon$ 
expansion \cite{15} and the functional renormalization group \cite{11}.}
\end{center}
\end{table}}

\sect{XY model exponents.}

Having established the critical exponents for the non-abelian 
Nambu-Jona-Lasinio model, (\ref{lagnjluniv}), we can now derive those for the 
chiral XY model in the abelian limit. It is clear that there are no singular 
colour group factors for the results of (\ref{njleta1}), (\ref{njlchi1}), 
(\ref{njleta2}), (\ref{chisig2}), (\ref{chipi2}), (\ref{njllam1}), 
(\ref{njllam2}) and (\ref{njleta3}). Therefore it is safe to take the abelian 
limit of (\ref{xylim}) giving
\begin{eqnarray}
\eta_1^{\mbox{\footnotesize{XY}}} &=&
-~ \frac{4\Gamma(2\mu-1)}{\mu\Gamma(1-\mu)\Gamma(\mu-1)\Gamma^2(\mu)}
\nonumber \\
\chi_{\sigma\,1}^{\mbox{\footnotesize{XY}}} &=& 
\chi_{\pi\,1}^{\mbox{\footnotesize{XY}}} ~=~ 0
\nonumber \\
\lambda_1^{\mbox{\footnotesize{XY}}} &=& -~ (2\mu-1) 
\eta_1^{\mbox{\footnotesize{XY}}}
\label{expxy1}
\end{eqnarray}
at leading order which clearly illustrates the issue surrounding the vanishing 
leading order vertex anomalous dimension for both Yukawa interactions. At next 
order we have 
\begin{eqnarray}
\eta_2^{\mbox{\footnotesize{XY}}} &=& \left[ \frac{(5\mu-1)}{2\mu(\mu-1)}
+ \Psi(\mu) \right] {\eta_1^{\mbox{\footnotesize{XY}}}}^2 
\nonumber \\
\chi_{\sigma\,2}^{\mbox{\footnotesize{XY}}} &=& 
\chi_{\pi\,2}^{\mbox{\footnotesize{XY}}} ~=~
-~ \frac{\mu^2(4\mu^2-10\mu+7)}{2(\mu-1)^3}
{\eta_1^{\mbox{\footnotesize{XY}}}}^2 
\nonumber \\
\lambda_2^{\mbox{\footnotesize{XY}}} &=& 
\left[ \frac{8\mu(\mu-1)}{(\mu-2)^2 \eta_1^{\mbox{\footnotesize{XY}}}} 
+ \frac{(8\mu^6-52\mu^5+108\mu^4-68\mu^3-21\mu^2+24\mu-4)}{2\mu(\mu-1)(\mu-2)^2}
\right. \nonumber \\
&& \left. ~
- \frac{2(2\mu^6-13\mu^5+31\mu^4-32\mu^3+9\mu^2+6\mu-2)}{(\mu-1)^2(\mu-2)^2}
\Psi(\mu)
\right. \nonumber \\
&& \left. ~
- \frac{\mu^2(2\mu-3)}{(\mu-1)(\mu-2)} \left[ \Phi(\mu) + \Psi^2(\mu) \right]
+ \frac{3\mu^2(3\mu-4)}{2(\mu-1)(\mu-2)} \Theta(\mu) \right]
{\eta_1^{\mbox{\footnotesize{XY}}}}^2 ~.
\label{expxy2}
\end{eqnarray}
We have checked that these expressions are in agreement with earlier work,
\cite{17,24}. Finally we obtain the main result for this model which is
\begin{eqnarray}
\eta_3^{\mbox{\footnotesize{XY}}} &=& \left[
\frac{1}{2} \left[ \Phi(\mu) + 3 \Psi^2(\mu) \right]
- \frac{(2\mu^4-3\mu^3-15\mu^2+18\mu-3)}{2\mu(\mu-1)^2} \Psi(\mu)
- \frac{1}{4}\Theta(\mu)
\right. \nonumber \\
&& \left. ~
- \frac{(4\mu^7-4\mu^6-7\mu^5-26\mu^4+84\mu^3-68\mu^2+20\mu-2)}{4\mu^2(\mu-1)^4}
\right] {\eta_1^{\mbox{\footnotesize{XY}}}}^3 ~. 
\label{expxyeta3}
\end{eqnarray}
The expression is considerably simpler than its non-abelian counterpart similar
to the $O(1/N^2)$ exponents. One aspect of this is that the function $\Xi(\mu)$
is absent which is not unrelated to the vanishing of $\chi_\sigma$ and 
$\chi_\pi$ at leading order.

{\begin{figure}[ht]
\includegraphics[width=7cm,height=7cm]{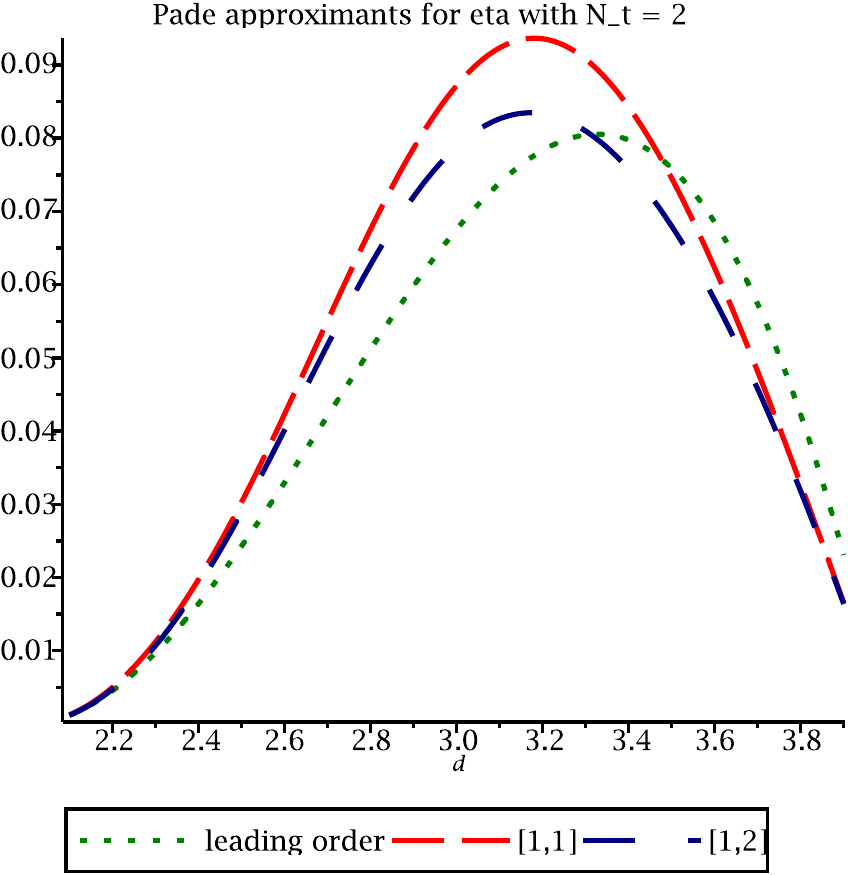}
\quad \quad \quad
\includegraphics[width=7cm,height=7cm]{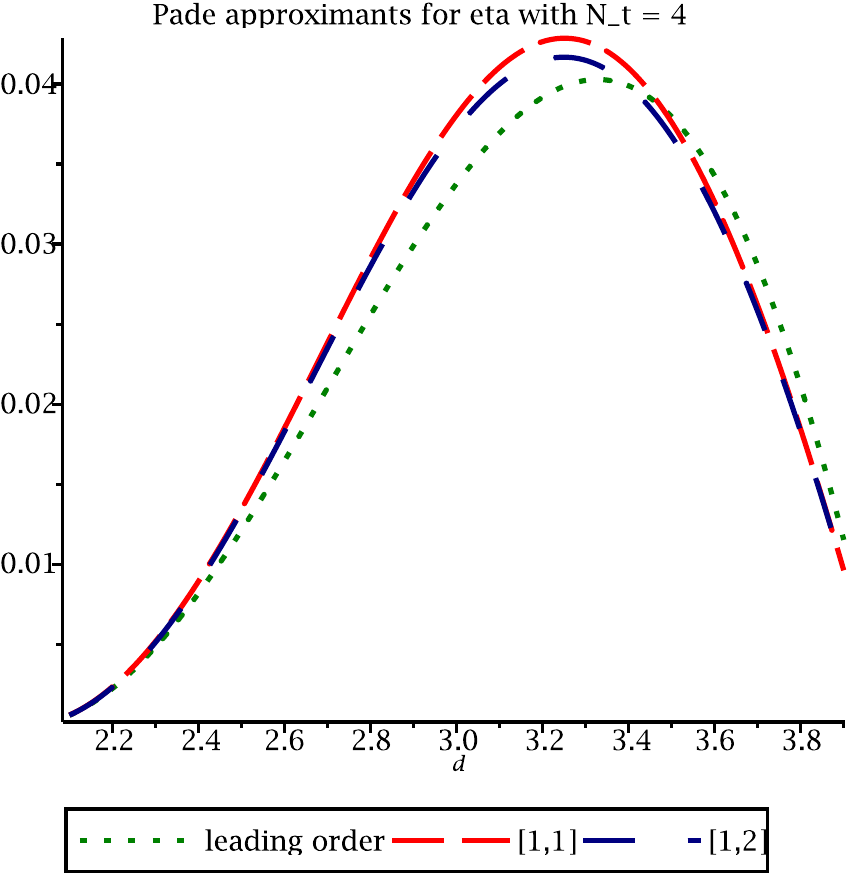}

\vspace{0.5cm}
\includegraphics[width=7cm,height=7cm]{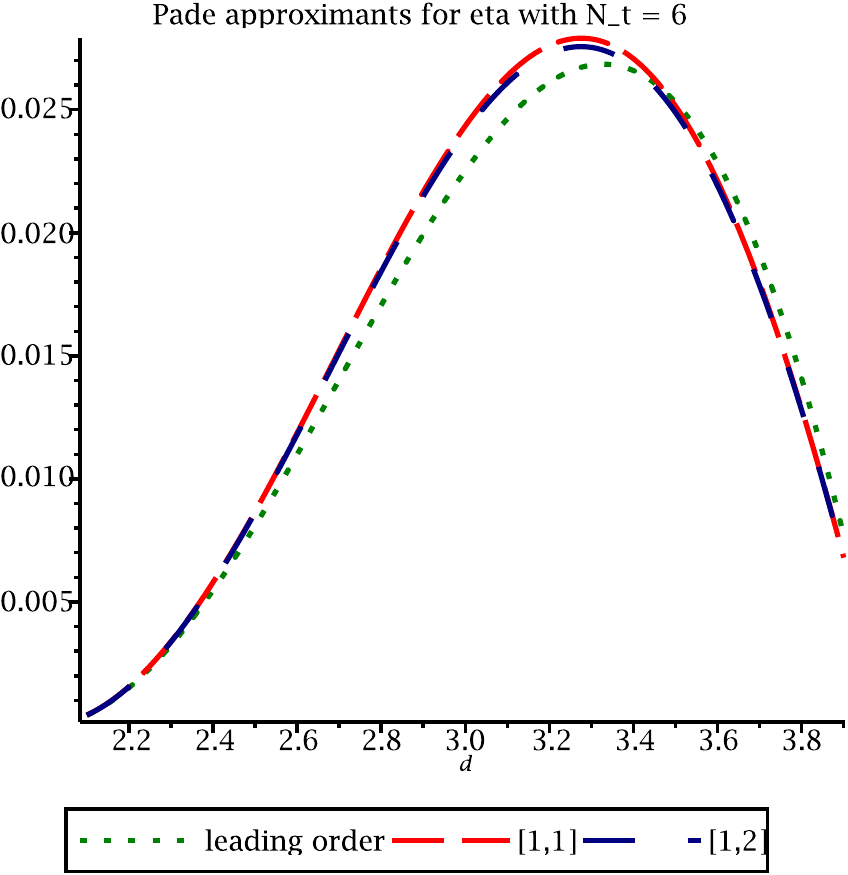}
\quad \quad \quad
\includegraphics[width=7cm,height=7cm]{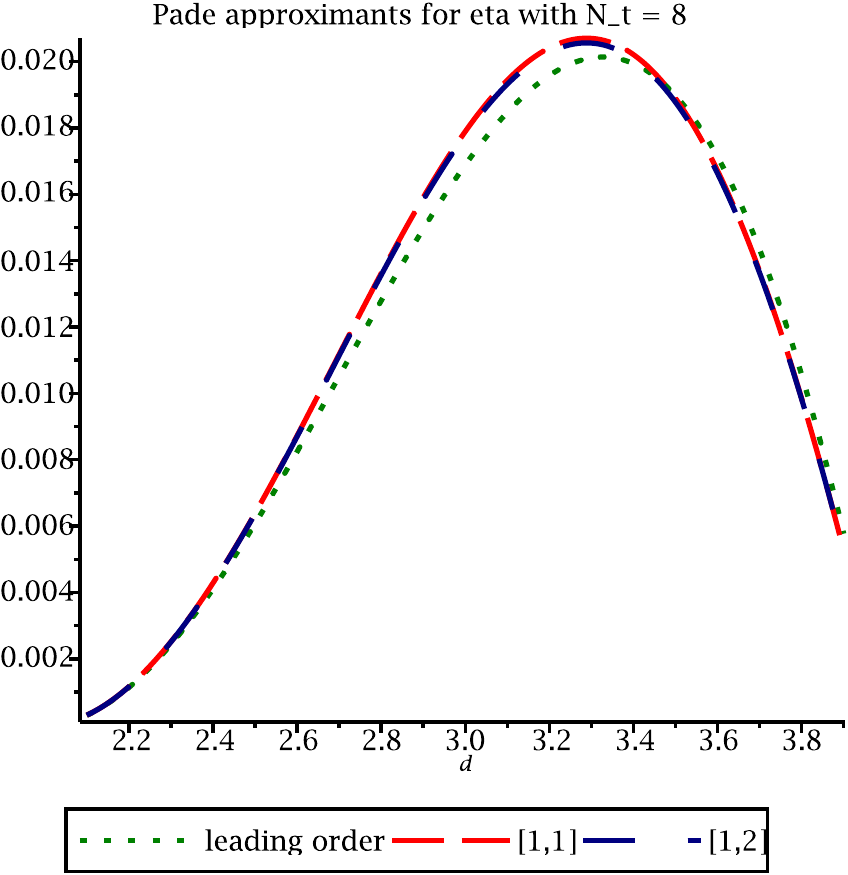}

\caption{Pad\'{e} approximants of $\eta^{\mbox{\footnotesize{XY}}}$ for
$N_t$~$=$~$2$, $4$, $6$ and $8$.}
\label{figxyeta}
\end{figure}}

While the exponents to $O(1/N^2)$ agree with previous derivations, it is
possible to check them against perturbative computations in four dimensions.
In \cite{15} the $\beta$-function and anomalous dimensions of the chiral XY
model were determined to four loops in the $\MSbar$ scheme. Moreover the 
$\epsilon$ expansion of the corresponding critical exponents were computed to
$O(\epsilon^4)$. Therefore if we expand our expressions in the same $\epsilon$
expansion they ought to be in agreement. Setting $d$~$=$~$4$~$-$~$2\epsilon$ in 
(\ref{expxy1}), (\ref{expxy2}) and (\ref{expxyeta3}) we find 
\begin{eqnarray}
\left. \eta^{\mbox{\footnotesize{XY}}} \right|_{d=4-2\epsilon} &=&
\left[ \epsilon - \frac{3}{2} \epsilon^2 - \frac{3}{4} \epsilon^3
+ \left[ 2 \zeta_3 - \frac{3}{8} \right] \epsilon^4 + O(\epsilon^5)
\right] \frac{1}{N_t} \nonumber \\
&& +~ \left[ -~ \epsilon + \frac{23}{4} \epsilon^2 - \frac{43}{8} \epsilon^3
- \left[ 8 \zeta_3 + \frac{19}{4} \right] \epsilon^4 + O(\epsilon^5)
\right] \frac{1}{N_t^2} \nonumber \\
&& +~ \left[ \epsilon - \frac{43}{4} \epsilon^2 + \frac{77}{4} \epsilon^3
+ \left[ \frac{39}{2} \zeta_3 + \frac{581}{16} \right] \epsilon^4 
+ O(\epsilon^5) \right] \frac{1}{N_t^3} ~+~ 
O \left( \frac{1}{N_t^4} \right) \nonumber \\
\left. \eta_\phi^{\mbox{\footnotesize{XY}}} \right|_{d=4-2\epsilon} &=&
2\epsilon 
+ \left[ -~ 2 \epsilon + 3 \epsilon^2 + \frac{3}{2} \epsilon^3
+ \left[ \frac{3}{4} - 4 \zeta_3 \right] \epsilon^4 + O(\epsilon^5)
\right] \frac{1}{N_t} \nonumber \\
&& + \left[ 2 \epsilon + \frac{1}{2} \epsilon^2 - \frac{101}{4} \epsilon^3
+ \left[ 16 \zeta_3 + \frac{51}{2} \right] \epsilon^4 + O(\epsilon^5)
\right] \frac{1}{N_t^2} ~+~
O \left( \frac{1}{N_t^3} \right) \nonumber \\
\left. \frac{1}{\nu^{\mbox{\footnotesize{XY}}}} \right|_{d=4-2\epsilon} &=&
2 - 2\epsilon 
+ \left[ -~ 6 \epsilon + 13 \epsilon^2 - \frac{3}{2} \epsilon^3
- \left[ \frac{3}{4} + 12 \zeta_3 \right] \epsilon^4 + O(\epsilon^5)
\right] \frac{1}{N_t} \nonumber \\
&& + \left[ 78 \epsilon - \frac{497}{2} \epsilon^2 
+ \left[ \frac{453}{4} - 120 \zeta_3 \right] \epsilon^3
\right. \nonumber \\
&& \left. ~~~
+ \left[ 240 \zeta_5 - 180 \zeta_4 + 744 \zeta_3 + \frac{205}{2} \right] 
\epsilon^4 + O(\epsilon^5)
\right] \frac{1}{N_t^2} ~+~
O \left( \frac{1}{N_t^3} \right) 
\end{eqnarray}
where the combination
\begin{equation}
\eta_\phi ~=~ 4 ~-~ 2 \mu ~-~ 2 \eta ~-~ 2 \chi_\sigma
\end{equation}
tallies with the definition used in \cite{15,16}. We have also set 
$N$~$=$~$4N_t$ to adjust for the different trace and group conventions used 
here in the underlying master integrals for the skeleton Dyson-Schwinger and 
conformal bootstrap equations. So $N_t$ corresponds to the $N$ used in 
\cite{15,16}. It is straightforward to verify that each expansion is in full 
agreement with \cite{15} after allowing for a different convention in the 
definition of $d$ in terms of $\epsilon$. In light of this agreement with high 
order perturbation theory we can obtain expressions for the exponents in three 
dimensions. Analytically we have  
\begin{eqnarray}
\left. \eta^{\mbox{\footnotesize{XY}}} \right|_{d=3} &=&
\frac{4}{3\pi^2 N_t} ~+~
\frac{160}{27\pi^4 N_t^2} ~-~
\frac{80 [ 3 \pi^2 + 20 ]}{243\pi^6 N_t^3} ~+~
O \left( \frac{1}{N_t^4} \right) \nonumber \\
\left. \eta_\phi^{\mbox{\footnotesize{XY}}} \right|_{d=3} &=&
1 ~-~ \frac{8}{3\pi^2 N_t} ~+~ \frac{544}{27\pi^4 N_t^2} ~+~
O \left( \frac{1}{N_t^3} \right) \nonumber \\
\left. \frac{1}{\nu^{\mbox{\footnotesize{XY}}}} \right|_{d=3} &=&
1 ~-~ \frac{16}{3\pi^2 N_t} ~+~ \frac{[ 216 \pi^2 + 2912 ]}{27\pi^6 N_t^2} ~+~
O \left( \frac{1}{N_t^3} \right) 
\end{eqnarray}
or 
\begin{eqnarray}
\left. \eta^{\mbox{\footnotesize{XY}}} \right|_{d=3} &=&
0.135095 \frac{1}{N_t} ~+~ 0.060835 \frac{1}{N_t^2} ~-~
0.016988 \frac{1}{N_t^3} ~+~
O \left( \frac{1}{N_t^4} \right) \nonumber \\
\left. \eta_\phi^{\mbox{\footnotesize{XY}}} \right|_{d=3} &=&
1 ~-~ 0.270190 \frac{1}{N_t} ~+~ 0.206840 \frac{1}{N_t^2} ~+~
O \left( \frac{1}{N_t^3} \right) \nonumber \\
\left. \frac{1}{\nu^{\mbox{\footnotesize{XY}}}} \right|_{d=3} &=&
1 ~-~ 0.540380\frac{1}{N_t} ~+~ 1.917775 \frac{1}{N_t^2} ~+~
O \left( \frac{1}{N_t^3} \right)
\end{eqnarray}
numerically. 

{\begin{figure}[ht]
\includegraphics[width=7cm,height=7cm]{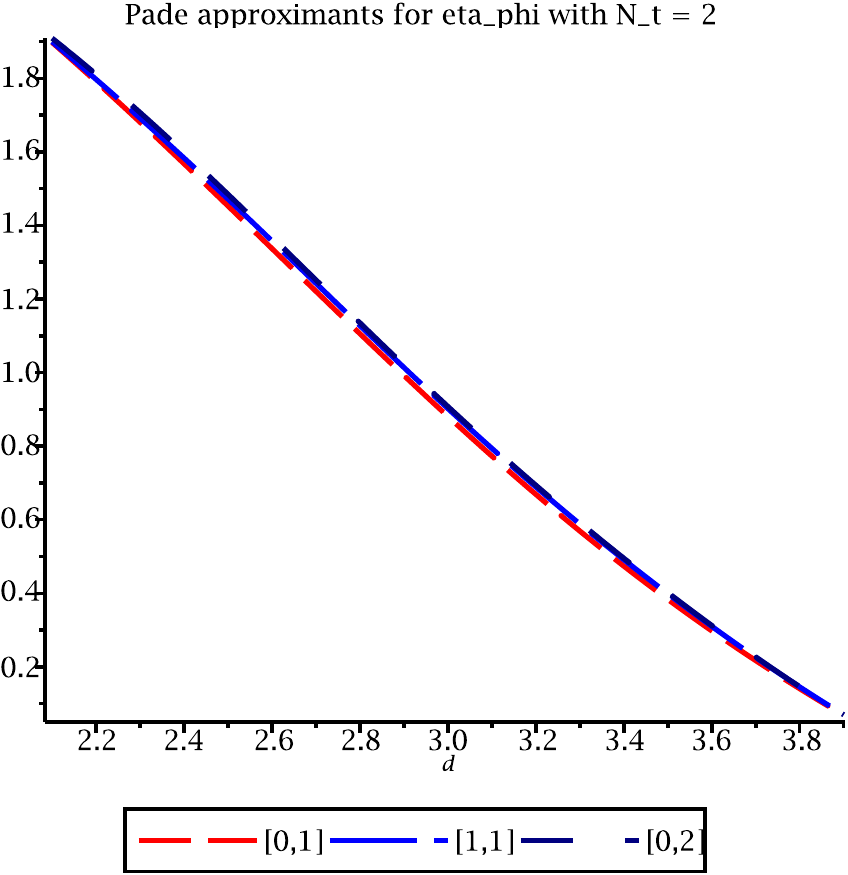}
\quad \quad \quad
\includegraphics[width=7cm,height=7cm]{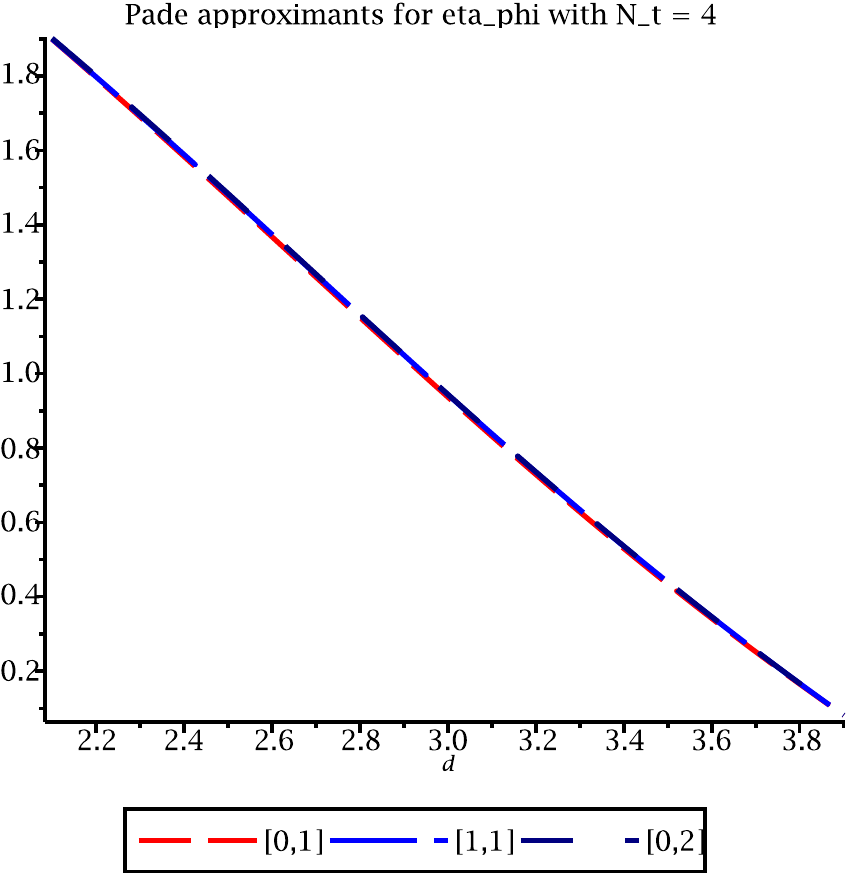}

\vspace{0.5cm}
\includegraphics[width=7cm,height=7cm]{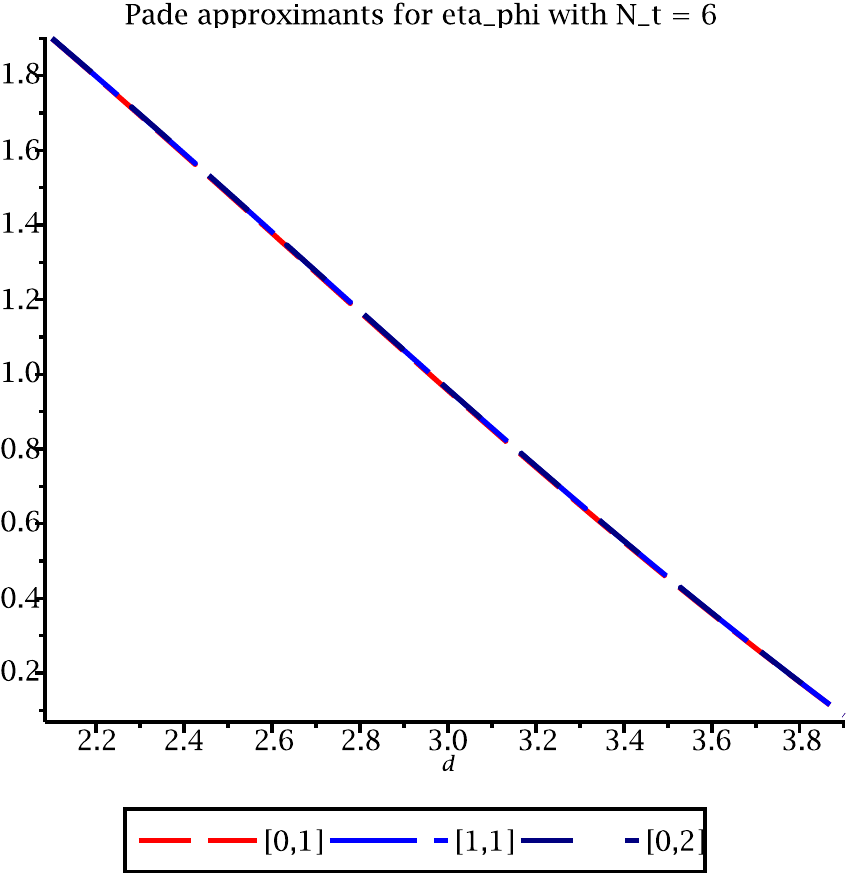}
\quad \quad \quad
\includegraphics[width=7cm,height=7cm]{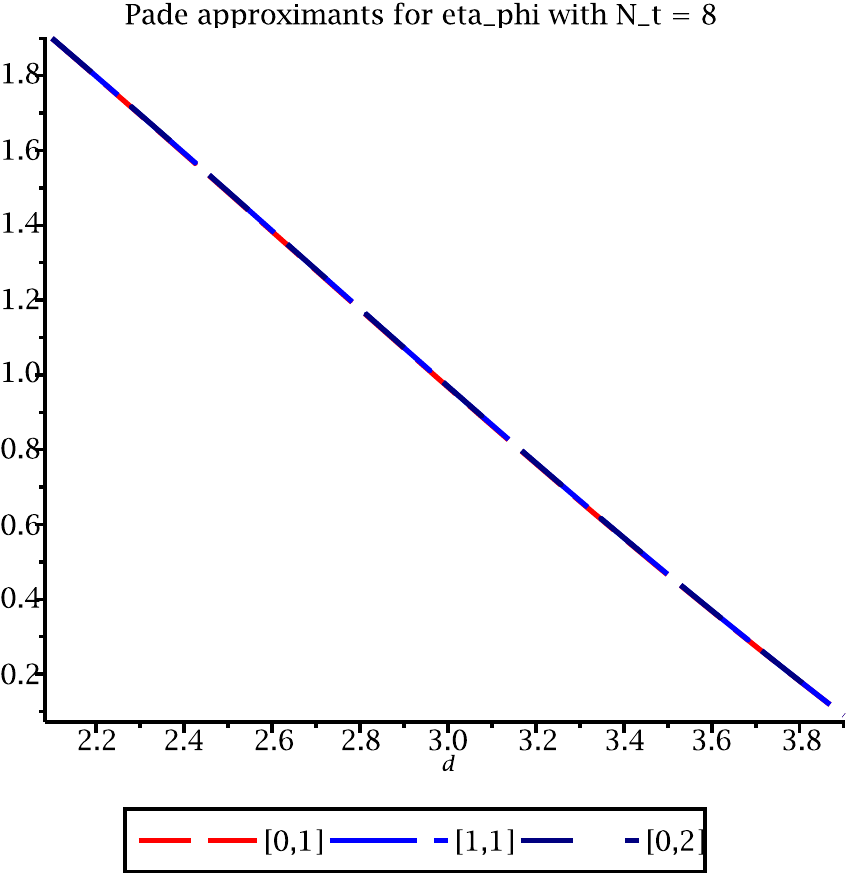}

\caption{Pad\'{e} approximants of $\eta_\phi^{\mbox{\footnotesize{XY}}}$ for
$N_t$~$=$~$2$, $4$, $6$ and $8$.}
\label{figxyetaphi}
\end{figure}}

The next stage is to extract estimates for the three dimensional exponents for
the case of interest in \cite{15} which corresponds to $N_t$~$=$~$2$ here. To
improve the convergence of the expansion we have followed a similar approach
to \cite{28} and evaluated several Pad\'{e} approximants for each of the
exponents. These have been plotted in Figures \ref{figxyeta}, \ref{figxyetaphi}
and \ref{figxynuinv} for $N_t$~$=$~$2$, $4$, $6$ and $8$ for 
$2$~$<$~$d$~$<$~$4$ for $\eta^{\mbox{\footnotesize{XY}}}$,
$\eta_\phi^{\mbox{\footnotesize{XY}}}$ and $1/\nu^{\mbox{\footnotesize{XY}}}$ 
respectively with numerical values for three dimensions recorded in Table $1$. 
The latter three values of $N_t$ are selected to gauge the effect the higher 
order $1/N$ corrections have. In the case of $\eta$ it is worth noting that we 
can plot the behaviour of $\eta_3$ unlike \cite{28}. This is because $\Xi(\mu)$
is absent from $\eta_3$ and it was not possible to write the function in an 
analytic form that could be passed to the plotting routine in that case. For 
$\eta^{\mbox{\footnotesize{XY}}}$ the $[2,1]$ Pad\'{e} had a singularity at 
around $d$~$=$~$3.6$. Equally for $N_t$~$=$~$2$ the $[0,2]$ Pad\'{e} of 
$1/\nu^{\mbox{\footnotesize{XY}}}$ had a maximum near a similar value of $d$. 
We have omitted plots of both Pad\'{e}'s for each exponent.  

What is evident in the plots of $\eta^{\mbox{\footnotesize{XY}}}$ for various
$N_t$ is that there is convergence of the $O(1/N^3)$ values down to 
$N_t$~$=$~$4$. For $N_t$~$=$~$2$ there is a difference between the $O(1/N^2)$
and $O(1/N^3)$ curves but the latter is bounded by the leading two orders. This
is similar to higher values of $N_t$. Given this it might not be unreasonable
to give $0.083$ as a rough estimate of $\eta^{\mbox{\footnotesize{XY}}}$ for
$N_t$~$=$~$2$ being the average of the two Pad\'{e} estimates. This is not
inconsistent with the $N_t$~$=$~$2$ estimates from other approaches which are 
summarized in Table $2$. In the case of $\eta_\phi^{\mbox{\footnotesize{XY}}}$ 
it is clear from Figure \ref{figxyetaphi} that there is very little difference 
for the value of the exponent between successive orders in $1/N$. Indeed 
comparing the values for $N_t$~$=$~$2$ with other estimates there is a degree 
of consistency. Finally the situation for $1/\nu^{\mbox{\footnotesize{XY}}}$ 
does not appear to be as good especially at leading order for low values of 
$N_t$ due to the kink that is evident in Figure \ref{figxynuinv} in its 
behaviour across $d$-dimensions. Although it appears that this feature washes 
out for larger values of $N_t$, it is not clear whether this would be the case 
for $N_t$~$=$~$2$ if an $O(1/N^3)$ expression were available. Again one could 
assume that the two large $N$ Pad\'{e} approximants for 
$1/\nu^{\mbox{\footnotesize{XY}}}$ were bounds of the true result. In this 
instance one would arrive at an estimate of around $0.845$ which is not 
dissimilar to the perturbative Pad\'{e} or functional renormalization group 
estimates of \cite{11,15}. 

{\begin{figure}[ht]
\includegraphics[width=7cm,height=7cm]{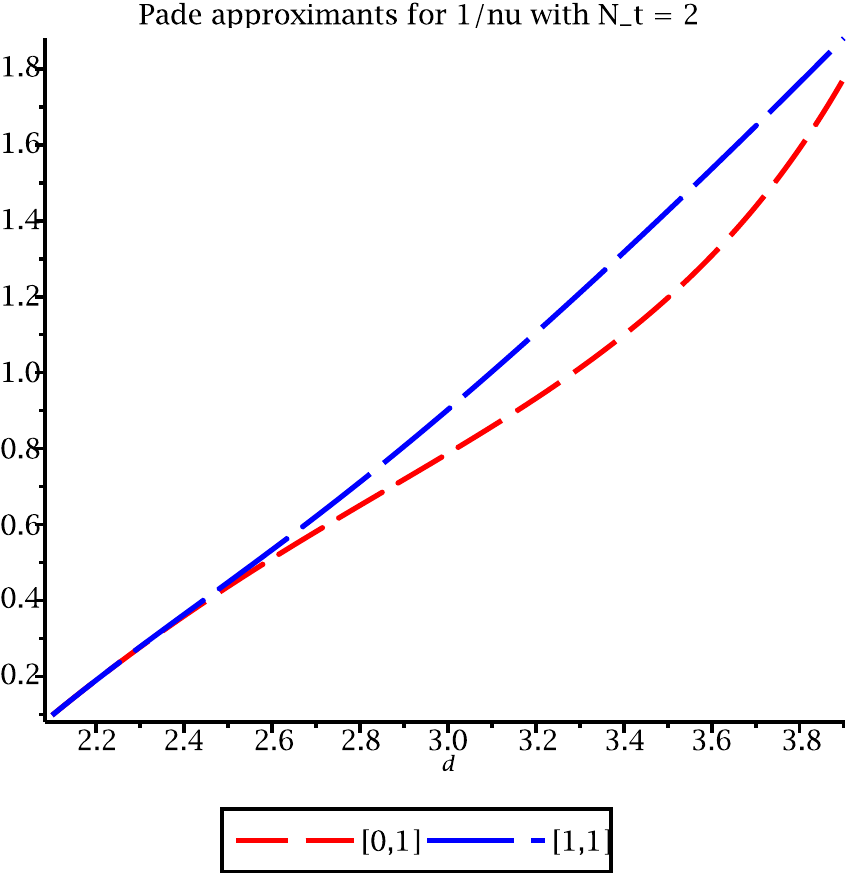}
\quad \quad \quad
\includegraphics[width=7cm,height=7cm]{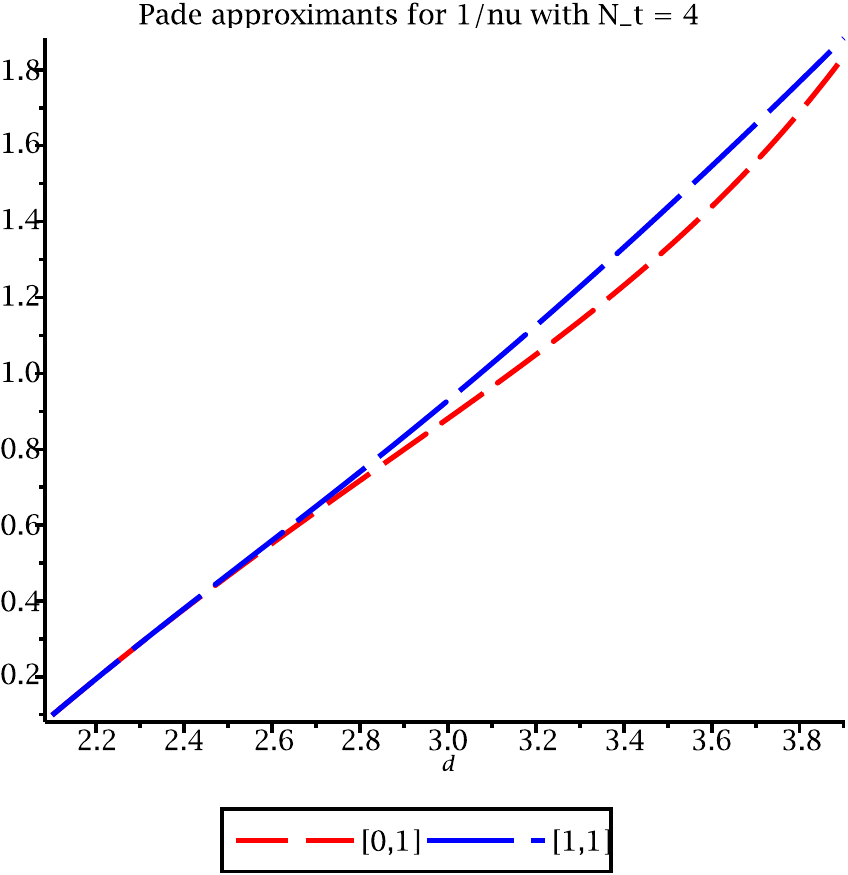}

\vspace{0.5cm}
\includegraphics[width=7cm,height=7cm]{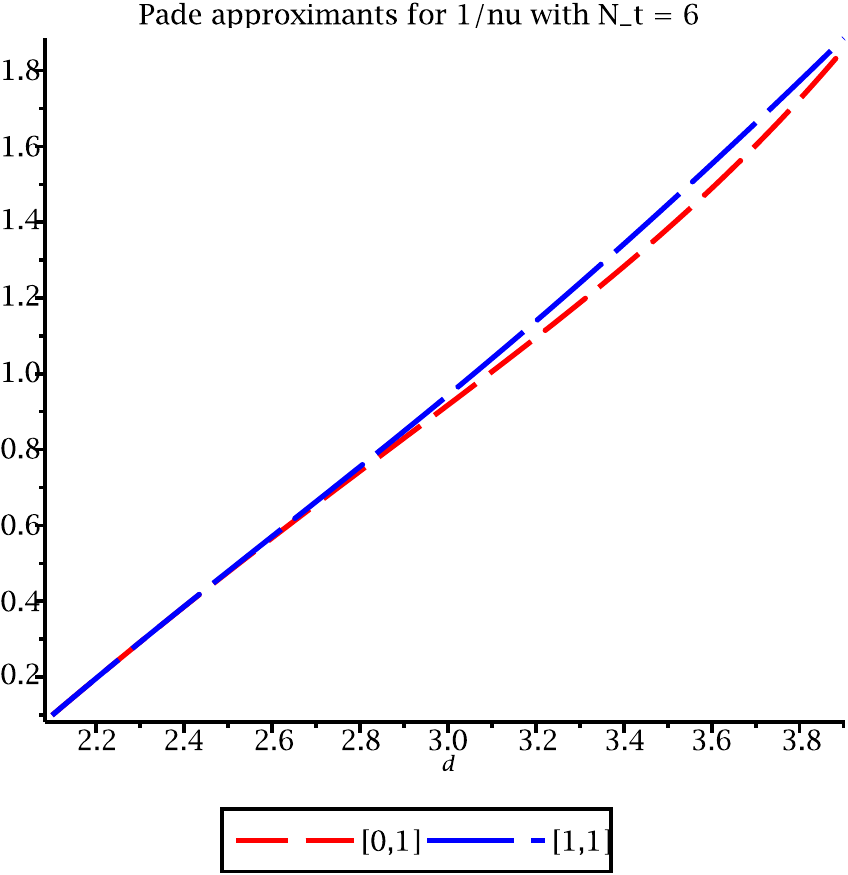}
\quad \quad \quad
\includegraphics[width=7cm,height=7cm]{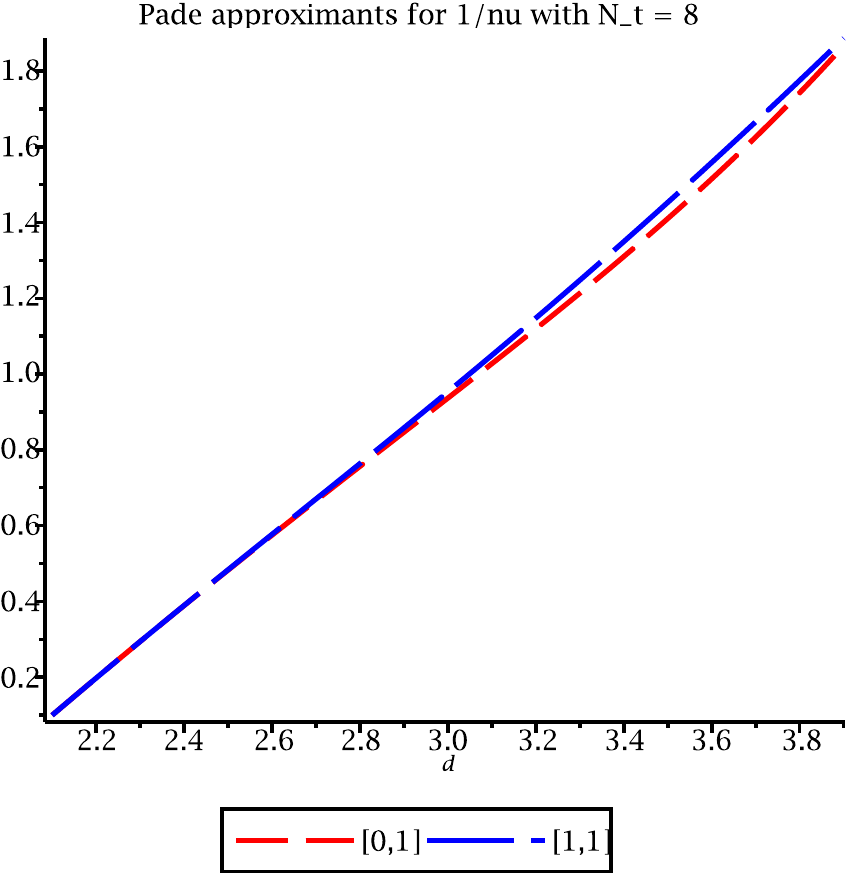}

\caption{Pad\'{e} approximants of $\frac{1}{\nu^{\mbox{\footnotesize{XY}}}}$ 
for $N_t$~$=$~$2$, $4$, $6$ and $8$.}
\label{figxynuinv}
\end{figure}}

\sect{Discussion.}

We have provided the large $N$ critical exponents for the Nambu-Jona-Lasinio
universality class with Lagrangian (\ref{lagnjluniv}) for a general Lie group 
using the large $N$ critical point formalism pioneered in \cite{18,19,20}. This
included the large $N$ conformal bootstrap formalism that allowed us to extract
$\eta_3$. To achieve this we had to extend the formalism to the situation where
there are two independent $3$-point interactions. By independent we mean the 
vertex anomalous dimensions of the separate vertices were inequivalent. The 
derivation of the underlying bootstrap equations was straightforward but was
discussed at length. Indeed from that it is evident that they can be readily 
generalized to the case of more independent $3$-point interactions. Taking the 
more general approach with a general Lie group means that results for exponents
in various models can be extracted in certain limits. Previously $\eta_3$ was 
calculated for a limited number of fermionic models including the $SU(2)$ 
Nambu-Jona-Lasinio model, \cite{35}, as the anomalous dimensions of each vertex
were the same. To produce the general Lie group results we were able to benefit
from the {\sc Form} encoded {\tt color.h} package which was based on \cite{42} 
that allowed us to express our exponents in terms of the general group Casimir 
invariants. This was particularly important for the light-by-light graphs that 
contribute to several exponents at $O(1/N^2)$ and $\eta_3$. One of the main 
consequences was that the abelian limit could be taken smoothly to deduce 
exponents for the chiral XY model. This circumvented the application of the 
bootstrap formalism directly to (\ref{lagxyuniv}) due to an ill-defined 
intermediate quantity that formally obstructs the direct derivation in the 
abelian case. The ultimate general expressions are analytic in the colour group
invariants allowing the chiral XY model exponents to emerge smoothly. Finally
this more general group approach for the large $N$ formalism offers a more
compact way of extracting critical exponents for other universality classes.
One benefit for instance is that it gives access to additional information on 
the terms in the explicit perturbative series of the underlying renormalization
group functions. 

\vspace{1cm}
\noindent
{\bf Acknowledgements.} This work was supported by a DFG Mercator Fellowship
and in part through the STFC Consolidated ST/T000988/1. The graphs were drawn
with the {\sc Axodraw} package \cite{49}. Computations were carried out using 
the symbolic manipulation language {\sc Form}, \cite{40,41}.

\end{document}